\documentclass[aps,pra,twocolumn,groupedaddress,nobibnotes]{revtex4-1}  
\usepackage{dcolumn,graphicx,amsmath,amssymb,algorithm,algpseudocode}
\usepackage{todonotes}
\usepackage{qcircuit}
\usepackage[colorlinks=true,linkcolor=blue,citecolor=blue,urlcolor=blue]{hyperref}
\usepackage[capitalise]{cleveref}
\usepackage{orcidlink}

\newcommand{\br}{\mathbf{r}}

\newcommand{\bR}{\mathbf{R}}

\begin{document}

\definecolor{brickred}{rgb}{.72,0,0} 


\title{
Towards the Simulation of Large Scale Protein-Ligand Interactions on NISQ-era Quantum Computers
}

\author{Fionn D.~Malone~\orcidlink{0000-0001-9239-0162}}
\author{Robert M.~Parrish~\orcidlink{0000-0002-2406-4741}}
\email{rob.parrish@qcware.com}
\author{Alicia R.~Welden~\orcidlink{0000-0002-2238-9825}}
\affiliation{QC Ware Corporation, Palo Alto, CA, 94301, USA}
\author{Thomas Fox~\orcidlink{0000-0002-1054-4701}}
\affiliation{Medicinal Chemistry, Boehringer Ingelheim Pharma GmbH \& Co. KG, Birkendorfer Stra{\ss}e 65, 88397 Biberach an der Ri\ss, Germany}
\author{Matthias Degroote~\orcidlink{0000-0002-8850-7708}}
\author{Elica ~Kyoseva~\orcidlink{0000-0002-9154-0293}}
\author{Nikolaj Moll~\orcidlink{0000-0001-5645-4667}}
\email{nikolaj.moll@boehringer-ingelheim.com}
\author{Raffaele Santagati~\orcidlink{0000-0001-9645-0580}}
\author{Michael Streif~\orcidlink{0000-0002-7509-4748}}
\affiliation{Quantum Lab, Boehringer Ingelheim, 55218 Ingelheim am Rhein, Germany}

\begin{abstract}
We explore the use of symmetry-adapted perturbation theory (SAPT) as a simple
and efficient means to compute interaction energies between large molecular
systems with a hybrid method combing NISQ-era quantum and classical computers.
From the one- and two-particle reduced density matrices of the monomer wavefunctions
obtained by the variational quantum eigensolver (VQE), we compute
SAPT contributions to
the interaction energy [SAPT(VQE)]. At first order, this energy yields the electrostatic
and exchange contributions for non-covalently bound systems. We empirically
find from ideal statevector simulations that the SAPT(VQE) interaction energy
components display orders of magnitude lower absolute errors than the
corresponding VQE total energies.  Therefore, even with coarsely optimized
low-depth VQE wavefunctions, we still obtain sub kcal/mol accuracy in the SAPT
interaction energies. In SAPT(VQE), the quantum requirements, such as qubit
count and circuit depth, are lowered by performing computations on the separate
molecular systems. Furthermore, active spaces allow for large systems
containing thousands of orbitals to be reduced to a small enough orbital set to
perform the quantum portions of the computations. We benchmark SAPT(VQE) (with the VQE component simulated
by ideal state-vector simulators) against a handful of small multi-reference
dimer systems and the iron center containing human cancer-relevant protein
lysine-specific demethylase 5 (KDM5A).
\end{abstract}


\maketitle


Quantum chemistry has emerged as one of the most promising areas where a
practical quantum advantage from near term quantum computers could be
demonstrated \citep{Aspuru-Guzik2005,McArdleRev2020,Cao2019}.  Identifying industrially
relevant applications that can practically benefit from quantum simulations is,
however, a complicated task \citep{Bauer2020,vonBurgCatalysis2021}.  On the one hand, existing
classical algorithms have benefited from decades of development, benchmarking
and optimization so demonstrating a computational advantage over these is
challenging given limitations on current quantum hardware
\citep{Motta2017,Williams2020,Eriksen2020}. On the other hand, current noisy
intermediate-scale quantum (NISQ) hardware \citep{Preskill2018} suffers from
relatively poor gate fidelity so that the resulting physical properties can
often be biased and far from exact without error mitigation \citep{Cohn2021}.
Thus it is important to design quantum algorithms that minimize the quantum
resources required.

Coupled with these challenges is the problem of finding an industrially relevant
application that can benefit from a quantum computer in the first place
\citep{ElfvingRelevant2020}.  One such area that has been suggested as possibly
benefiting from quantum computing is computer aided drug design (CADD)
\citep{HeifetzCADD2020}.  CADD has a long history and has many components
ranging from high-level optimization problems such as structure search and
conformational sampling down to low-level quantum mechanical problems such as
computing protein-ligand interaction energies, all of which could potentially
benefit from a large scale quantum computer \citep{Cao2018,Outeiral2021}.  In
this work we will focus on this final problem, namely computing the interaction
energy and properties of large scale protein-ligand systems by approximately
solving the electronic structure problem.

To date, most quantum algorithms aimed at solving the electronic structure
problem directly are general and have mostly been applied to small
molecules in small basis sets.  Of course, applications are limited by current
quantum resources so that reaching chemical accuracy, defined as calculating
energy differences in the complete basis set limit to within 1 kcal/mol
accuracy, is difficult to achieve.  Nevertheless, relatively minor attention has
been paid to the entire workflow required to solve an industrially relevant
problem in drug design.  This often includes highly tailored approaches that
require classical pre- and post-processing, molecular dynamics and structure
relaxation, active space selection and finally computation of interaction
energies and related quantities all for systems containing hundreds or thousands
of atoms.  Thus, it is important to isolate potential application areas now
and codify these workflows with quantum algorithms in mind as the pace of hardware improvement accelerates.

Conceptually, computing the interaction energy of a protein-ligand system is a
straightforward task.  One computes the ground state energy of the dimer and
monomer systems separately and subtracts the two to determine the interaction
energy.  There are a number of issues with this approach, particularly when a
quantum computer is involved.  First, in finite Gaussian basis sets typically
employed in quantum chemical computations, one has to account for basis set
superposition error (BSSE) using the counterpoise correction \citep{BoysCP1970}.  This
unnecessarily increases the qubit count requirements for the individual monomers and
can potentially lead to convergence issues for hybrid quantum-classical
algorithms like the variational quantum eigensolver (VQE) \citep{Peruzzo2014,McCleanVQE2016}.
A more concerning problem for NISQ computers is resolving total energies of
individual monomers (typically on the order of 1000s of kcal/mol) to
sufficient precision to subtractively resolve binding energies which are
typically on the order of 5 kcal/mol.  This is a major issue for NISQ
approaches which typically evaluate total energy expectation values
statistically, carrying very high measurement cost penalty for high-precision
expectation values.  This is also an increasingly challenging problem for
heuristic algorithms like the VQE which would require very deep circuits with
thousands of parameters to achieve the required precision (note that precision
and accuracy are the same concern in subtracting total energies unless strict
relative error cancellation can be ensured, which is not clear with methods like
VQE).
This may be practically impossible with the current general algorithms and
available hardware although we note alternatives to the VQE may help to overcome
this issue \citep{Huggins2021}.

In this work we propose using symmetry adapted perturbation theory (SAPT)
\citep{Jeziorski1976,Jeziorski1994} to directly compute the interaction energy
through direct expectation values rather than differences, which overcomes some
of these problems. Firstly, in principle SAPT does not suffer from BSSE as it
directly computes the interaction energy as a perturbation series in the
intermolecular potential (note that SAPT still suffers from basis set
incompleteness error, as with all second-quantized methods). Secondly, monomer-centered basis sets can be used which can afford additional savings if the
geometry of the monomers is fixed across the dissociation path, potentially
reducing the number of different quantum computations that have to be performed.
Moreover, as SAPT directly computes the interaction energy as a sum of
expectation values (rather than differences of large expectation values), it
often exhibits favorable error cancellation for errors inherent to the chosen
wavefunction ansatz. We show that this observation can significantly reduce the
resource requirements for circuit depth with only very coarse VQE wavefunctions
required for sub kcal/mol accuracy in the interaction energy components.
Finally, SAPT offers a physically motivated breakdown of the interaction energy
components into electrostatic, exchange, induction and dispersion contributions
which can offer valuable insight for medicinal chemists when designing
protein inhibitors \citep{Parrish2014}.

Beyond suggesting SAPT as a useful approach for NISQ quantum computers we outline an efficient active space formulation of SAPT that can be applied for protein-ligand interactions for systems containing heavy metal centers and thousands of atoms.
Key to this implementation is the GPU accelerated classical pre- and post-processing steps which practically help to run such simulations \citep{Parrish2018}.  We will largely focus on the accurate qualitative description of systems with strong multi-reference character that can not easily be described by classical approaches and thus could offer a more transparent demonstration of a practical quantum advantage.
In this paper we will only consider the first order contributions to the exchange energy leaving the second order terms, which require solving electronic response, for future work.

We begin by outlining in detail the active space formulation for first-order SAPT that can 
be coupled to any quantum simulation that can produce one- and two-particle reduced 
density matrices.
Next we discuss the VQE ansatz used in this work, although the SAPT method itself is 
largely independent of the way in which the ground state properties are computed.
Finally we benchmark our method using ideal quantum simulators and demonstrate a 
significant reduction in error found when poorly converged VQE wavefunctions are used for 
model multi-reference systems and for the human cancer relevant \citep{Yang2021} 
lysine-specific demethylase 5 (KDM5A) protein with different ligand substitutions.

\section{Methods}

In this section we will describe the classical implementation of the density matrix formulation of SAPT followed by how this implementation can be efficiently adapted for NISQ devices.

\subsection{Notation}

In this work we consider the interaction of two monomers $A$ and $B$ and will
use the following notation for different real orbital types belonging to $A$ and/or $B$
\begin{itemize}
\item $\mu/\nu$ - nonorthogonal atomic spatial orbital basis indices. Note that
these could conceptually be either monomer-centered of dimer-centered bases as
far as the theory is concerned. Unless otherwise noted, we use 
monomer-centered bases for all atomic spatial orbital bases encountered in
practical test cases in this work. 
\item $p/q$ - orthogonal molecular spatial orbital basis indices.
\item $i/j$ - orthogonal occupied spatial orbital basis indices.
\item $t/u$ - orthogonal active spatial orbital basis indices.
\item $a/b$ - orthogonal virtual spatial orbital basis indices.
\end{itemize}
Repeated indices within a monomer will be denoted with primes, e.g., $p, p',
p'', p'''$.
Summation over repeated indices is assumed throughout.
When dealing with spin-orbital quantities, we use the context
specific notation of an ``unbarred'' orbital index to denote $\alpha$ and a
``barred'' orbital index to denote $\beta$, i.e., $p^\dagger$ is an $\alpha$
spin-orbital creation operator on spatial orbital index $p$, while $\bar
p^\dagger$ is a $\beta$ spin-orbital creation operator on spatial orbital index
$p$.

\subsection{Symmetry Adapted Perturbation Theory}

Traditionally, the interaction energy between two monomers, $E_\mathrm{int}$, is calculated in the supermolecular approach as
\begin{equation}
    E_\mathrm{int} = E_{AB} - E_{A} - E_B,
\end{equation}
where $E_{AB}$ is the ground state energy of the combined system and $E_{A/B}$
are the energies of the individual fragments, evaluated at the frozen geometry
of the dimer system (i.e., no deformation energy contributions).
In contrast, in SAPT, $E_\mathrm{int}$ is instead evaluated through a perturbation series in the intermonomer interaction potential
\begin{align}
    V = \sum_{i=1}^{N_A}\sum_{j=1}^{N_B} \tilde{v}(\br_i,\br_j)\label{eq:potential},
\end{align}
where $N_{A/B}$ is the number of electrons in monomer $A$ or $B$, and
\begin{equation}
\begin{split}
\tilde{v}(\br_i,\br_j)
=
&\frac{1}{|\br_i-\br_j|}
\\
-
&\frac{1}{N_B}\sum_{J\in B}Z_J\frac{1}{|\bR_J-\br_i|}
-
\frac{1}{N_A}\sum_{I\in A}Z_I\frac{1}{|\bR_I-\br_j|}
\\
+
&\frac{1}{N_AN_B}\sum_{I\in A}\sum_{I\in B}Z_I Z_J\frac{1}{|\bR_I-\bR_J|},
\end{split}
\end{equation}
describes the interaction of the electrons and nuclei in monomer $A$ with the electrons and nuclei in monomer $B$ (and vice versa).
Symmetrized Rayleigh-Schr{\"o}dinger perturbation theory yields a perturbation series for the interaction energy
\begin{equation}
E_\mathrm{int}
=
E_\mathrm{pol}^{(1)}
+
E_\mathrm{exch}^{(1)}
+
E_\mathrm{pol}^{(2)}
+
E_\mathrm{exch}^{(2)}
+
\dots
\label{eq:sapt}
\end{equation}
where, $E_\mathrm{pol}^{(1)}$ is the electrostatic energy, $E_\mathrm{pol}^{(2)}$ is a sum of dispersion and induction energies, while $E_{\mathrm{exch}}^{(n)}$ account for exchange interactions.

In order to evaluate the first two terms in the perturbation series in \cref{eq:sapt} one first needs to solve for the ground state wavefunction, $|\Psi_A\rangle$, of the individual monomer Hamiltonian, here given for monomer $A$
\begin{equation}
\begin{split}
    \hat{H}_A = &\sum_{pp'} h_{pp'} \hat{E}_{pp'} + \\
                &\frac{1}{2}\sum_{pp'p''p'''} v^{pp'}_{p''p'''} (\hat E_{pp''}\hat E_{p'p'''}-\delta_{p''p'}\hat E_{pp'''}),
\end{split}
\label{eq:ham}
\end{equation}
where $h_{pq}$ and $v^{pp'}_{p''p'''}=(pp''|p'p''')$ are the usual one- and two-electron integrals and
\begin{equation}
    \hat E_{pp'} = p^\dagger p' + \bar p^\dagger \bar p',
    \label{eq:excitation}
\end{equation}
is the singlet-adapted one-particle substitution operator.

In the absence of the exact ground state monomer wavefunctions, the SAPT interaction energy is traditionally computed as a triple perturbation theory in the intramonomer fluctuation potentials (assuming a M{\o}ller-Plesset partitioning of the Hamiltonian).
At the lowest order this gives rise to SAPT(HF) (or SAPT0), where each term is evaluated using Hartree--Fock density matrices\citep{Jeziorski1994}.
Another popular approach is to use density functional theory SAPT(DFT) \citep{Williams2000,Misquitta2002} to account for intramonomer electron correlation, but this approach is naturally limited by the performance of the chosen density functional \citep{Hapka}.

In this work we will assume that we can determine the exact (or at least very accurate) ground state properties of the individual monomers using a quantum computer.
The first order SAPT expressions can then be evaluated using only the ground state unperturbed wavefunctions of the individual wavefunctions, i.e., $|\Psi^0\rangle = |\Psi_A\rangle \otimes|\Psi_B\rangle$.
We will use the density matrix formulation of SAPT \citep{Moszynski1994} systematized by Korona \citep{KoronaExchange2008,Korona2008,Korona2009} and recently fully implemented for complete active space self consistent field (CASSCF) wavefunctions by Hapka and {\it et al.} \citep{Hapka}
This formalism allows for the evaluation of the terms appearing in \cref{eq:sapt} using just the ground state one- and two-particle reduced density matrices of the monomers with additional response terms for the second order terms.
Detailed derivations of the density matrix formation of SAPT are given elsewhere and here we will focus on the efficient implementation in terms of optimized chemistry primitives on quantum computers.

\subsection{Density Matrix Formulation of SAPT}

The first order polarization energy is the electrostatic repulsion energy of monomer $A$ and $B$, and is given by
\begin{align}
E_{\mathrm{pol}}^{(1)}
&=
\langle \Psi_{A} \Psi_{B} |
\hat V
| \Psi_{A} \Psi_{B} \rangle
\\
&=
\bar{\gamma}_{pp'} \tilde{v}_{pq}^{p'q'} \bar{\gamma}_{qq'}
\label{eq:elst1}
\end{align}
where, $\bar{\gamma}_{pp'}$ is the spin-summed one-particle reduced density matrix
\begin{equation}
    \bar{\gamma}_{pp'} = \langle \Psi_A|p^\dagger p + \bar p^\dagger \bar p|\Psi_A \rangle,\label{eq:opdm}
\end{equation}
and we have introduced the generalized two-electron repulsion integrals
\begin{equation}
\begin{split}
\tilde{v}_{pq}^{p'q'}
=
&(pp'|V|qq')
\\
=
&(pp'|qq')
\\
+
&\frac{1}{N_A}(V_A|qq')S_{pp'}
+
\frac{1}{N_B}(pp'|V_B)S_{qq'}
\\
+
&\frac{V_{AB}}{N_AN_B}S_{pp'}S_{qq'},
\label{eq:gen_eri}
\end{split}
\end{equation}
where $(pp'|qq')$ is a mixed two-electron electron repulsion integral between orbitals on monomer $A$ and $B$, $S_{pp'} = (p|p')$ is an overlap integral and $(V_X|qq')$ are matrix elements of the electron-ion potential of monomer $X$ in the orbital basis of the other monomer.

Similarly, the first order exchange energy is given as \citep{Moszynski1994}
\begin{equation}
E_{\mathrm{exch}}^{(1)}
=
\frac{
    \langle \Psi_A \Psi_B |
    V \mathcal{A} |
    \Psi_A \Psi_B\rangle
}
{
    \langle \Psi_A \Psi_B |
    \mathcal{A} \Psi_A \Psi_B\rangle
}
\\
-E_{\mathrm{pol}}^{(1)}\label{eq:exch},
\end{equation}
where $\mathcal{A}$ is the antisymmetrizer operator.
Using the $S^2$ approximation, and the density matrix formulation of SAPT \citep{Moszynski1994}, it can be shown that \cref{eq:exch} can be written as
\begin{align}
E_{\mathrm{exch}}^{(1)}
&=
-\frac{1}{2}(\bar{\gamma}_{pp'} \bar{\gamma}_{qq'} \tilde{v}_{pq}^{q'p'} \\
\begin{split}
&+\bar{\gamma}_{pp'}
\bar{\Gamma}_{q''q'''}^{qq'}
S_{p'q'}
\tilde{v}_{pq}^{q''' q''}\\
&+\bar{\gamma}_{qq'}
\bar{\Gamma}_{p''p'''}^{pp'}
S_{q'p'}
\tilde{v}_{pq}^{p'' p'''}\\
&\bar{\Gamma}_{p''p'''}^{pp'}
\bar{\Gamma}_{q''q'''}^{qq'}
S_{p'q'''}
S_{q'p'''}
\tilde{v}_{pq}^{p'' q''}\\
&-E_{\mathrm{pol}}^{(1)}S_{pq'} \gamma_{qq'} S_{qp'} \gamma_{pp'}
),
\end{split}\\
&= T_1 + T_2 + T_3 + T_4 + T_5
\label{eq:exch_dm}
\end{align}
where
\begin{equation}
    \bar{\Gamma}^{pp'}_{p''p'''} = 2 (\Gamma^{p p'}_{p'' p'''} + \Gamma_{p'' \bar p'''}^{p \bar p'}),
\end{equation}
is the spin-summed two-particle reduced density matrix,
where throughout this work we assume the monomers have singlet ground states, and
\begin{equation}
\Gamma^{p \bar p'}_{p'' \bar p'''}
=
\langle \Psi_A |
p^\dagger \bar p'^\dagger \bar p''' p''
|\Psi_A\rangle.\label{eq:tpdm}
\end{equation}
As can be seen from \cref{eq:elst1} and \cref{eq:exch_dm} the first order SAPT expressions require only the one- and two-particle density matrix to be evaluated.

\subsection{Efficient Active Space Implementation\label{sec:active_space}}

Given that NISQ-era devices are currently limited to tens of qubits (spin-orbitals) we will use an active space approach in order to tackle protein-ligand interactions.
We will heavily leverage standard quantum chemistry primitives such as integral driven Coulomb and exchange matrix builds which exploit sparsity \citep{Almlof1982} and can efficiently be implemented on GPUs \citep{UfimtsevGPU12008,UfimtsevGPU22009,UfimtsevGPU32009}.
These considerations are important when simulating 100s of atoms and thousands of basis functions.

In the active space approach we partition the one-electron orbital set into
$N_c$ core orbitals, $N_a$ active orbitals and $N_i$ virtual orbitals.
This partitioning gives rise to modified monomer Hamiltonians given by (for example for monomer $A$)
\begin{equation}
    \begin{split}
    \hat{H_{A}}' =
    &\sum_{tt'\sigma} \tilde{h}_{tt'} a^\dagger_{t\sigma}a_{t'\sigma}+\\
    &\sum_{\sigma\sigma'}\sum_{tt't''t'''} (tt''|t't''') a^\dagger_{t\sigma}a^\dagger_{t'\sigma'} a_{t'''\sigma'} a_{t''\sigma}
    \end{split}\label{eq:fzc},
\end{equation}
where the modified one-electron integrals $\tilde{h}_{tt'}$ now include core-active space interactions
\begin{equation}
    \tilde{h}_{tt'} = h_{tt'} + \sum_{ii'} \left[(tt'|ii') - \frac{1}{2} (ti'|it')\right]\gamma_{ii'}\label{eq:core_ham}.
\end{equation}
For large scale applications $N_c$ is typically large (100-1000) and \cref{eq:core_ham} can be efficiently evaluated in the AO basis
\begin{equation}
    \tilde{h}_{\mu\mu'} = h_{\mu\mu'} + 2 J^c_{\mu\mu'}-K^c_{\mu\mu'},\label{eq:core_ao}
\end{equation}
before being transformed to the active space MO basis
\begin{equation}
    \tilde{h}_{tt'} = \sum_{\mu\mu'} C_{\mu t} \tilde{h}_{\mu\mu'} C_{\mu't'}
\end{equation}
where $C$ are the molecular orbital coefficients.
In \cref{eq:core_ao} we have introduced the usual (core) Coulomb and exchange matrices
\begin{align}
    J^c_{\mu\mu'} \equiv J[\gamma^c]_{\mu\mu'} &= \sum_{\mu''\mu'''} (\mu\mu'|\mu''\mu''')\gamma^c_{\mu''\mu'''}\label{eq:K}\\
    K^c_{\mu\mu'} \equiv K[\gamma^c]_{\mu\mu'} &= \sum_{\mu''\mu'''} (\mu\mu'''|\mu''\mu')\gamma^c_{\mu''\mu'''}\label{eq:J},
\end{align}
where the core density matrix is given by
\begin{equation}
    \gamma^c_{\mu\mu'} = \sum_i C_{\mu i} C_{\mu' i}.
\end{equation}

Once the ground state in \cref{eq:fzc} has been found and the active space one- and two-particle density matrices have been formed the first order SAPT contributions can be calculated as a classical post processing step which can be implemented efficiently by considering the block structure of the one- and two-particle density matrices.
Recall that in the MO basis we have
\begin{align}
    \bar{\gamma}^c = \bar{\gamma}_{ii'}  &= 2\delta_{ii'}\\
      \bar{\gamma}^a = \gamma_{tt'} &= \gamma_{tt'}
\end{align}
and
\begin{align}
    \bar{\Gamma}^{cc} &= \bar{\Gamma}^{ii'}_{i''i'''} = \bar{\gamma}_{ii''}\bar{\gamma}_{ii'''}-\frac{1}{2}\bar{\gamma}_{ii'''}\bar{\gamma}_{ii''}\\
    \bar{\Gamma}^{ac} & =
    \begin{cases}
        \bar{\Gamma}^{ti}_{t'i'} &= \bar{\Gamma}^{it}_{i't'} = \bar{\gamma}_{tt'}\bar{\gamma}_{ii'}\\
        \bar{\Gamma}^{it}_{t'i} &= \bar{\Gamma}^{ti}_{it'} = -\frac{1}{2}\bar{\gamma}_{tt'}\bar{\gamma}_{ii'} \\
    \end{cases}\\
    \bar{\Gamma}^{aa} &= \bar{\Gamma}^{tt'}_{t''t'''} = \bar{\Gamma}^{tt'}_{t''t'''},
    \label{eq:2pdm_ac}
\end{align}
where again $i$ and $t$ are occupied (core) and active orbital indices respectively.
All other blocks of the density matrices are zero.

Not much is gained by exploiting this block structure for first order terms that contain the one-particle density matrix only.
Thus, we evaluate them in the AO basis directly as
\begin{equation}
    E_{\mathrm{pol}}^{(1)} = \sum_{\mu\mu'} \bar{\gamma}_{\mu\mu'} \tilde{J}[\bar{\gamma}^B]_{\mu\mu'},
\end{equation}
where the generalized Coulomb ($\tilde{J}$) and exchange ($\tilde{K}$) matrices  are analogues of \cref{eq:J,eq:K} with the standard electron repulsion integrals replaced with their generalized counterparts (see \cref{eq:gen_eri}) and the AO density matrices are given by
\begin{equation}
\bar{\gamma}_{\mu\mu'}
=
\sum_{ii'} C_{\mu i} \bar{\gamma}^c_{ii'} C_{\mu' i'}
+
\sum_{tt'} C_{\mu t} \bar{\gamma}^a_{tt'} C_{\mu' t'}.
\end{equation}
Similarly, the first term in the exchange matrix is evaluated as
\begin{equation}
    T_1 =
    -\frac{1}{2}\sum_{\mu\mu'}\bar{\gamma}_{\mu\mu'}
    \tilde{K}[\bar{\gamma}^B]_{\mu\mu'},
\end{equation}
while $T_5$ is a simple trace of a matrix product.
The other three terms are a bit more complicated.
For example, for $T_2$, there will be in total six terms arising from the different combinations of core and active orbitals sets.
Inserting the expressions for $\gamma^{c}$ and $\Gamma^{cc},\Gamma^{ac}$ from \cref{eq:2pdm_ac} we have
\begin{align}
\begin{split}
T_2[\bar \gamma_A^c, \bar \Gamma_B^{ac}]
=
4 \bar{\gamma}_{uu'}
&\left[
S_{ij}
(
\tilde{v}_{iu}^{ju' }
-
\frac{1}{2} \tilde{v}_{iu}^{u'j}
)
\right.
\\
+
&
\left.
S_{iu}
(\tilde{v}_{ij}^{u'j}
-
\frac{1}{2}\tilde{v}_{ij}^{ju'}
)
\right].
\end{split}
\end{align}
This can be simplified by using the definitions of the generalized Coulomb and exchange matrices defined earlier to be written as
\begin{equation}
\begin{split}
&T_2[\bar \gamma_A^c, \bar \Gamma_B^{ac}]
=
\\
&4 S_{ij} C_{\mu i} (\tilde{J}[\bar{\gamma}_B^{a}]_{\mu\nu''}-\frac{1}{2}\tilde{K}[\bar{\gamma}_B^a]_{\mu\nu''}) C_{\nu''j}\\
+
&2 \gamma_{uu'} S_{iu} C_{\mu i} (\tilde{J}[\bar{\gamma}_B^c]_{\mu\nu''}-\frac{1}{2}\tilde{K}[\bar{\gamma}_B^c]_{\mu\nu''}) C_{\nu'' u'}.
\end{split}
\end{equation}
Similarly we have
\begin{align}
T_2[\bar \gamma_A^a, \bar \Gamma_B^{cc}]
=
&2\gamma_{tt'}
S_{t'j'}
C_{\mu t}
\left(
    \tilde{J}[\gamma_B^{c}]_{\mu\nu'''}
-
\frac{1}{2}\tilde{K}[\gamma_B^{c}]_{\mu\nu'''}
\right)
C_{\nu''' j'}
,
\end{align}
and
\begin{equation}
\begin{split}
&T_2[\bar \gamma_A^a, \bar \Gamma_B^{ac}]
=
\\
&2 \bar{\gamma}_{tt'} S_{t'j} C_{\mu t} (\tilde{J}[\bar{\gamma}^a_B]_{\mu\nu''}-\frac{1}{2}\tilde{K}[\bar{\gamma}_B^a]_{\mu\nu''}) C_{\nu'' j}\\
+
&\bar{\gamma}_{tt'} \bar{\gamma}_{uu'} S_{t'u} C_{\mu t} (\tilde{J}[\bar{\gamma}^c_B]_{\mu\nu''}-\frac{1}{2}\tilde{K}[\bar{\gamma}^c_{B}]_{\mu\nu''}) C_{\nu''u'}.
\end{split}
\end{equation}
Terms involving $\Gamma^{aa}$ typically cannot be simplified much, e.g.,
\begin{equation}
    T_2[\bar \gamma_A^c, \bar \Gamma_B^{aa}] = 2 \bar{\Gamma}^{uu'}_{u''u'''} S_{iu'} \tilde{v}_{iu}^{u''' u''}.
\end{equation}
To avoid the formation of the $O(N_{a}^3 N_{c})$ generalized electron repulsion integrals we first form
\begin{equation}
    E_{\mu u'} = \sum_{i} C_{\mu i} S_{iu'}
\end{equation}
before forming the generalized MO-ERIs, so that
\begin{equation*}
    T_2[\bar \gamma_A^c, \bar \Gamma_B^{aa}] = 2 \bar{\Gamma}^{uu'}_{u''u'''} \tilde{v}_{u'u}^{u''' u''}.
\end{equation*}
Finally, we have
\begin{equation*}
    T_2[\bar \gamma_A^a, \bar \Gamma_B^{aa}]
=
\bar{\gamma}_{tt'}
\bar{\Gamma}^{uu'}_{u''u'''} S_{t'u'} \tilde{v}_{tu}^{u'''u''},
\end{equation*}
which does not simplify further.
Note that the generalized two-electron integrals only need to be constructed for the last two terms and require at most $\mathcal{O}(N_{\mathrm{act}}^4)$ storage.
Although not a concern for the system sizes considered here, further reduction in computational cost and memory can be achieved through density fitting and related approaches \citep{HohensteinDFSAPT02010} .

$T_3$ is analogous to $T_2$ while $T_4$ is quite verbose and contains sixteen terms. Full expressions for these are given in \cref{app:active}.
The above expressions are completely general and do not depend on the method used to evaluate the one- and two-particle density matrices.

\subsection{Variational Quantum Eigensolver\label{sec:VQE}}

Up to this point we have assumed that the ground state one- and two-particle reduced density matrices of the monomers could be determined by some means.
In this subsection we will give further details of the VQE algorithm used in
this work. As described, SAPT is essentially a post processing step that relies
only on the availability of the one- and two-particle reduced density matrices.
Therefore, it is not tied to any particular quantum algorithm, however, in this work we will focus on using the VQE.

In the SAPT(VQE) approach, one or both of the monomer active space
wavefunctions are generated by VQE-type quantum circuits
\begin{equation}
|\Psi_A\rangle
\equiv
|\Omega_A\rangle
|\Phi^c_{A}\rangle,
\end{equation}
where $|\Omega_{A}\rangle$ is the active space
wavefunction generated by a quantum circuit $\hat U_{\mathrm{VQE}}$ from a
starting state $|\Phi^0_{A}\rangle$ on a $2N_c$-qubit quantum computer, i.e.,
\begin{equation}
|\Omega_{A}\rangle
\equiv
\hat U_{\mathrm{VQE}}
|\Phi^0_{A}\rangle
\end{equation}
and $|\Phi^c_{A}\rangle$ is the closed-shell core determinant.
Note that with the VQE ansatz adopted for this paper, the active space
wavefunction $|\Omega_{A}\rangle$ will be taken to be real, and will be a definite
eigenfunction of the $\hat N_{\alpha}$, $\hat N_{\beta}$, and $\hat S^2$
operators.

In the Jordan-Wigner representation used in this paper, the
creation/annihilation operators are defined as,
\begin{equation}
p^{\pm}
=
\bigotimes_{p' = 0}^{p' = p-1}
\hat Z_{p'}
(\hat X_{p} \mp i\hat Y_{p}) / 2,
\end{equation}
\begin{equation}
\bar p^{\pm}
=
\bigotimes_{p' = 0}^{p' = N_\alpha - 1}
\hat Z_{p'}
\bigotimes_{\bar p' = 0}^{\bar p' = \bar p-1}
\hat Z_{\bar p'}
(\hat X_{\bar p} \mp i\hat Y_{\bar p}) / 2,
\end{equation}
where $p^+ = p^\dagger$ and $p^-=p$ and we order the Jordan-Wigner strings in
$\alpha$-then-$\beta$ order.

In this work we use a modified version of the unitary cluster Jastrow wavefunction
\citep{Matsuzawa2020} ($k$-uCJ) which takes the form
\begin{equation}
| \Psi_{0} \rangle
=
\prod_{k}
\exp(-\hat K^{(k)})
\exp(\hat T^{(k)})
\exp(+\hat K^{(k)})
| \Phi_{0} \rangle,
\label{eq:ansatz}
\end{equation}
where $\hat{K}^{(k)}$ and $\hat{T}^{(k)}$ are one- and two-body operators, and $k$ is a parameter that controls the depth of the circuit and as a result its variational freedom.
Our modified $k$-uCJ ansatz differs from Ref.~\citenum{Matsuzawa2020} in the choice of two-body operator and we restrict ourselves to real anti-symmetric matrices.
For the one-body rotations we use spin-restricted orbital transformations,
\begin{equation}
    \hat K^{(k)}
\equiv
\sum_{pp'}
\kappa_{pp'}^{(k)}
\left [
(p^\dagger p' - p'^\dagger p)
+
(\bar p^\dagger \bar p' - \bar p'^\dagger \bar p)
\right ]
\end{equation}
where $\kappa_{pp'}^{(k)} = -\kappa_{p'p}^{(k)}$ is a real, antisymmetric $M \times M$
matrix of orbital rotation generators. The restricted orbital transformation
operator is equivalent to a 1-particle spin-restricted orbital transformation
via,
\begin{equation}
    U_{pp'}^{(k)}
\equiv
\left [
    \exp (\kappa^{(k)})
\right ]_{pp'}
\end{equation}
Note that $\hat U^{(k)} \in \mathcal{SO}(M)$. This spin-restricted orbital rotation
can be efficiently implemented in quantum circuits via a fabric of Givens
rotations \citep{KivlichanGivens2018}.

For the two-particle operator we use a modified diagonal double-substitution operator,
\begin{equation}
\begin{split}
\hat T^{(k)}
\equiv
\sum_{p=0}^{M-1}
\sum_{\substack{p'=p\ \mathrm{mod} \ 2 \\p+=2}}^{M-2}
\tau_{pp'}^{(k)}
[
&
(p'+1)^{\dagger}
\overline{(p'+1)}^{\dagger}
p'
\bar p'
\\
&
-
p'^{\dagger}
\bar p'^{\dagger}
{(p'+1)}
\overline{(p'+1)}
]
\end{split}
\label{eq:diag_doub}
\end{equation}
which is similar to the unitary pair coupled-cluster generalized singles and doubles expression ($k$-UpCCGSD) \citep{LeeUCCSD2018}.
We note that normally a pair coupled-cluster approximation for the two-body operator would render $k$-uCJ equivalent to $k$-UpCCGSD \citep{Matsuzawa2020}, 
however the variant used here has the advantage that it avoids non-local Jordan-Wigner strings and thus is closer to the quantum number preserving fabric circuit \citep{AnselmettiLocal2021} or the fermionic SWAP network variant of $k$-UpCCGSD \citep{OGormanSWAP2019}.
However, as our choice is not quite any of these ansatzes, for the remainder of this work we will name it $k$-muCJ for clarity, with the `m' standing for modified.
We stress again that the choice of VQE ansatz is largely irrelevant from a SAPT perspective and is not a major point in this paper.

\begin{figure*}
\Qcircuit @C=1em @R=.7em {
& \gate{G} \qwx[2]  & \qw  & \qw  & \qw & \gate{G} \qwx[2]  & \qw  & \qw  & \qw & \multigate{3}{P_X}  & \qw  & \qw  & \qw & \multigate{3}{P_X}  & \qw  & \qw  & \qw & \gate{G} \qwx[2]  & \qw  & \qw  & \qw & \gate{G} \qwx[2]  & \qw  & \qw  & \qw  & \\
 & \qw & \gate{G} \qwx[2]  & \qw  & \qw  & \qw & \gate{G} \qwx[2]  & \qw  & \qw & \ghost{P_X}  & \qw  & \qw  & \qw & \ghost{P_X}  & \qw  & \qw  & \qw  & \qw & \gate{G} \qwx[2]  & \qw  & \qw  & \qw & \gate{G} \qwx[2]  & \qw  & \qw  & \\
& \gate{G}  & \qw & \gate{G} \qwx[2]  & \qw & \gate{G}  & \qw & \gate{G} \qwx[2]  & \qw & \ghost{P_X} & \multigate{3}{P_X}  & \qw  & \qw & \ghost{P_X} & \multigate{3}{P_X}  & \qw  & \qw & \gate{G}  & \qw & \gate{G} \qwx[2]  & \qw & \gate{G}  & \qw & \gate{G} \qwx[2]  & \qw  & \\
 & \qw & \gate{G}  & \qw & \gate{G} \qwx[2]  & \qw & \gate{G}  & \qw & \gate{G} \qwx[2] & \ghost{P_X} & \ghost{P_X}  & \qw  & \qw & \ghost{P_X} & \ghost{P_X}  & \qw  & \qw  & \qw & \gate{G}  & \qw & \gate{G} \qwx[2]  & \qw & \gate{G}  & \qw & \gate{G} \qwx[2]  & \\
& \gate{G} \qwx[2]  & \qw & \gate{G}  & \qw & \gate{G} \qwx[2]  & \qw & \gate{G}  & \qw & \multigate{3}{P_X} & \ghost{P_X}  & \qw  & \qw & \multigate{3}{P_X} & \ghost{P_X}  & \qw  & \qw & \gate{G} \qwx[2]  & \qw & \gate{G}  & \qw & \gate{G} \qwx[2]  & \qw & \gate{G}  & \qw  & \\
 & \qw & \gate{G} \qwx[2]  & \qw & \gate{G}  & \qw & \gate{G} \qwx[2]  & \qw & \gate{G} & \ghost{P_X} & \ghost{P_X}  & \qw  & \qw & \ghost{P_X} & \ghost{P_X}  & \qw  & \qw  & \qw & \gate{G} \qwx[2]  & \qw & \gate{G}  & \qw & \gate{G} \qwx[2]  & \qw & \gate{G}  & \\
& \gate{G}  & \qw  & \qw  & \qw & \gate{G}  & \qw  & \qw  & \qw & \ghost{P_X}  & \qw  & \qw  & \qw & \ghost{P_X}  & \qw  & \qw  & \qw & \gate{G}  & \qw  & \qw  & \qw & \gate{G}  & \qw  & \qw  & \qw  & \\
 & \qw & \gate{G}  & \qw  & \qw  & \qw & \gate{G}  & \qw  & \qw & \ghost{P_X}  & \qw  & \qw  & \qw & \ghost{P_X}  & \qw  & \qw  & \qw  & \qw & \gate{G}  & \qw  & \qw  & \qw & \gate{G}  & \qw  & \qw  & \\
}
\caption{Single layer ($k=1$) $k$-muCJ VQE entangler circuit used in this work sketched for $M=4$ spatial orbitals or $N=8$ qubits. Even (odd) qubits label $\alpha$ ($\beta$) spin-orbitals. The circuit consists of a single layer of orbital rotations which are implemented as a series of two-qubit Givens rotations, followed by a layer of diagonal double substituion operators implemented via the four-qubit pair-exchange gate given in \cref{eq:px_gate}, followed by another layer of orbital rotations. Note that for the spin restricted ansatz used here the angles in the Givens gate for $\alpha$ and $\beta$ spin-orbitals is the same.\label{fig:entangler}}
\end{figure*}
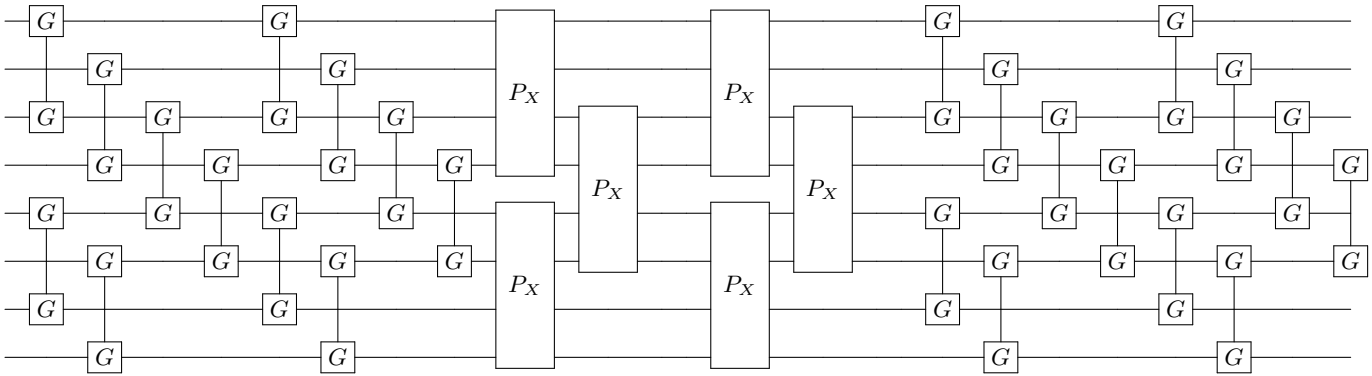

The diagonal doubles operator in \cref{eq:diag_doub} can be implemented as a product of four-qubit pair-exchange gates, $\hat{P}_X(\theta)$, which have the action,
\begin{equation}
    \hat {P}_X (\theta)
\equiv
\left [
\begin{array}{cccccccccccccccc}
1  & & & & & & & & & & & & & & & \\
 & 1 & & & & & & & & & & & & & & \\
 & & 1 & & & & & & & & & & & & & \\
 & & & c & & & & & & & & & -s & & & \\
 & & & & 1 & & & & & & & & & & & \\
 & & & & & 1 & & & & & & & & & & \\
 & & & & & & 1 & & & & & & & & & \\
 & & & & & & & 1 & & & & & & & & \\
 & & & & & & & & 1 & & & & & & & \\
 & & & & & & & & & 1 & & & & & & \\
 & & & & & & & & & & 1 & & & & & \\
 & & & & & & & & & & & 1 & & & & \\
 & & & +s & & & & & & & & & c & & & \\
 & & & & & & & & & & & & & 1 & & \\
 & & & & & & & & & & & & & & 1 & \\
 & & & & & & & & & & & & & & & 1 \\
\end{array}
\right ]
\label{eq:px_gate}
\end{equation}
in the four-qubit Hilbert space.
Note that,
\begin{align}
|3 \rangle
=
| 0011 \rangle
\ \mathrm{and},\
|12\rangle
=
| 1100\rangle,
\end{align}
and recalling that $\alpha$ and $\beta$ qubits are ordered in an interleaved fashion, we see the operator implements a Givens rotation implementing a partial double-substitution
in the closed-shell space.
The decomposition of \cref{eq:px_gate} into more standard gates is given in Ref.~\citenum{AnselmettiLocal2021}.
An example of one layer of the muCJ circuit ansatz is given in \cref{fig:entangler}.

With an ansatz of the form of \cref{eq:ansatz} we can write the VQE objective function as
\begin{align}
E_{0} 
(\kappa_{pq}^{k}, \tau_{pq}^{k})
&\equiv
\langle \Psi_{0} 
(\kappa_{pq}^{k}, \tau_{pq}^{k})
|
\hat H
|
\Psi_{0}
(\kappa_{pq}^{k}, \tau_{pq}^{k})
\rangle
\\
&=
\langle \Phi_{0} 
|
\hat U^{\dagger}
(\kappa_{pq}^{k}, \tau_{pq}^{k})
\hat H
\hat U 
(\kappa_{pq}^{k}, \tau_{pq}^{k})
|
\Phi_{0}
\rangle.
\label{eq:obj}
\end{align}
The VQE algorithm then proceeds in hybrid form by using the quantum computer to evaluate \cref{eq:obj} before the variational parameters $\{\kappa^k_{pq},\tau_{pq}^k\}$ are updated using a classical optimization algorithm.

To estimate the one- and two-particle reduced density matrices we write the expectation values in \cref{eq:opdm} and \cref{eq:tpdm} in terms of the Jordan-Wigner strings.
The number of measurements scales like $\mathcal{O}(N_a^4)$ which can be reduced to an extent by accounting for symmetries in the one- and two-particle density matrices.
Although not a focus of this paper, efficiently estimating density matrices is an active area of research and techniques exist to increase the parallelism of the measurements \citep{AIQuantum2020} and for error mitigation \citep{TillyCASSCF2021}.

\section{Results}

\subsection{Computational Details}

For the idealized experiments presented here we use the L-BFGS-B algorithm provided by scipy \citep{virtanen2020scipy} to optimize the VQE objective function and
used the RHF state as the initial wavefunction.
Classical SCF computations and integral generation was performed using Terachem \citep{UfimtsevGPU12008,UfimtsevGPU22009,UfimtsevGPU32009} interfaced through Lightspeed \citep{lightspeed}.
The GPU accelerated ideal VQE simulations were implemented using the quasar/vulcan codes.
Double factorization was used for evaluating the total energy of the VQE ansatz \citep{MottaDF2021,HugginsDF2021}.
VMD \citep{Humphrey1996} was used for visualizing molecular orbitals and molecular structures with the exception of the KDM5A system which used MOE \citep{VilarMOE2008}.

\subsection{Multi Reference Benchmarks}

To assess the accuracy of SAPT(VQE) we will apply the method to investigate the intermolecular electrostatic and exchange energy to a selection of dimers from the S22 benchmark set \citep{Jurecka2006}.
We will modify these systems to ensure one of the monomers in question has some multi reference character that cannot be well described by SAPT(RHF).
In all examples we will benchmark our results against classical SAPT(CASCI) that employs the same active space as the SAPT(VQE) computations.

\begin{figure*}
    \includegraphics[width=\linewidth]{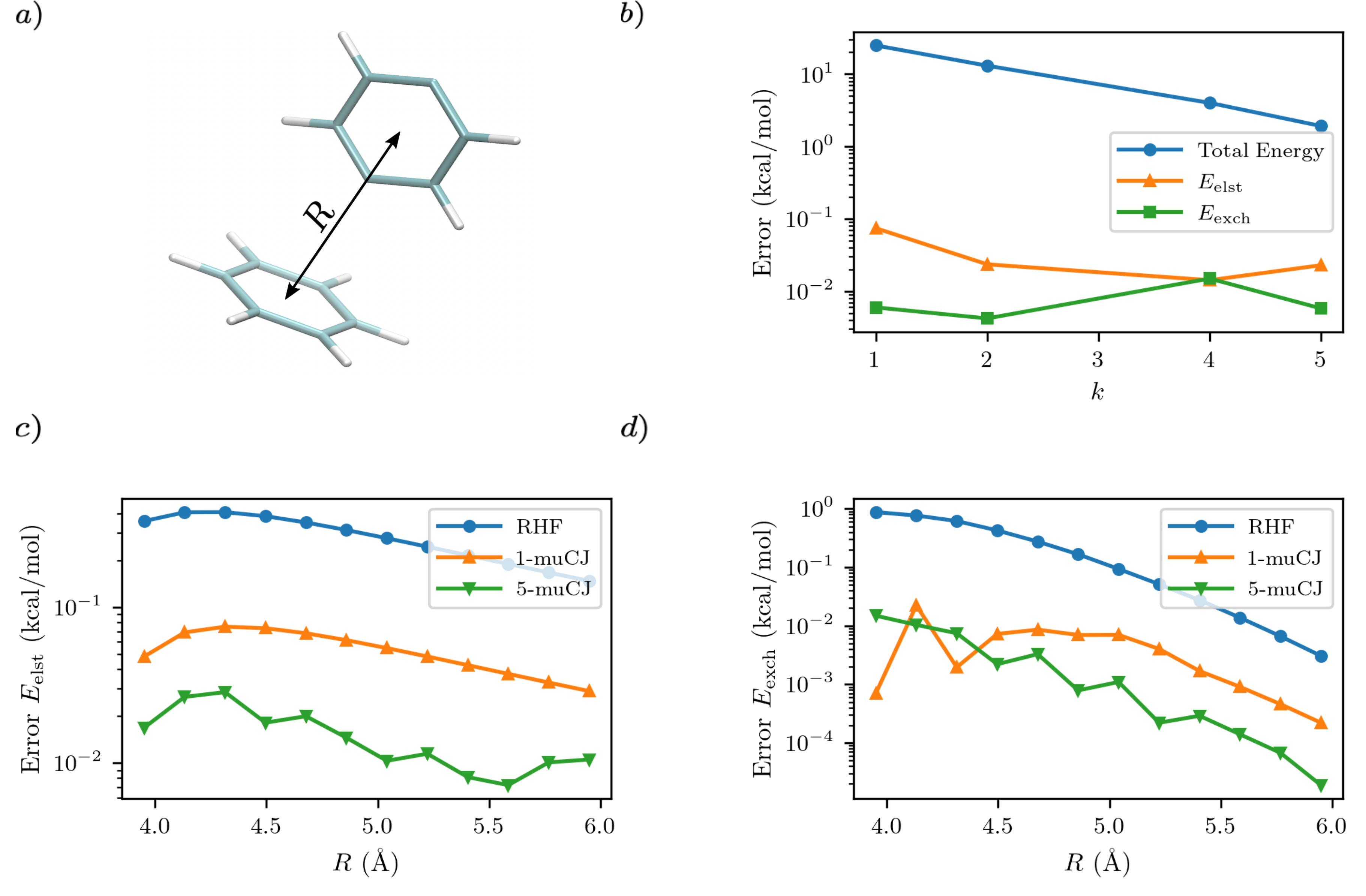}
\caption{(a) T-shaped configuration of benzene--p-benzyne dimer with the center-to-center intermonomer distance labelled as $R$. (b) Absolute error in p-benzyne monomer VQE total energy compared to errors in the SAPT(VQE) electrostatic and exchange energies at $R=4.45$ \AA \  intermonomer separation as a function of the circuit repetition factor $k$ in the $k$-muCJ ansatz. (c)-(d) Absolute errors relative to SAPT(CASCI) in the electrostatic and exchange energies calculated using different monomer wavefunctions for p-benzyne as a function of the intermonomer separation.\label{fig:benzene}}
\end{figure*}

For our first test case we will investigate a benzene--p-benzyne dimer arranged in the T-shape configuration visualized in \cref{fig:benzene} (a).
This is a variation on the classic benzene dimer SAPT benchmark, except p-benzyne has a biradical ground state and thus benefits from a multi-reference approach.
To describe p-benzyne we construct a (6e, 6o) active space from the HOMO-2 to LUMO+2 RHF/cc-pVDZ MOs and treat the benzene monomer at the RHF level of theory.
Previous results suggest that p-benzyne represents a challenge for VQE with the number of parameters required to reach chemical accuracy being roughly half the number of configuration state functions in the exact solution \citep{AnselmettiLocal2021}.
Thus, it represents an interesting test case to see how errors from VQE propagate through to the resultant SAPT energy components.

As can be seen from \cref{fig:benzene}, the p-benzyne molecule does represent a challenge for the $k$-muCJ ansatz, with a $k > 5$ required to reach an error in the total energy below 1 kcal/mol.
In contrast we see that the errors in the SAPT(VQE) electrostatic and exchange energies are 2-4 orders of magnitude smaller than the corresponding VQE total energy and that a shallow depth $k=1$ VQE ansatz would be sufficient for sub kcal/mol accuracy in these terms.
We see this holds across the dissociation curve, which is expected as the VQE solution is fixed when working in a monomer-centered basis set, which is another advantage of SAPT(VQE).
It should be noted that the SAPT(RHF) binding energy for this dimer system is on the order of 2 kcal/mol for this basis set.

\begin{figure*}
    \hspace{3em}
    \includegraphics[width=\linewidth]{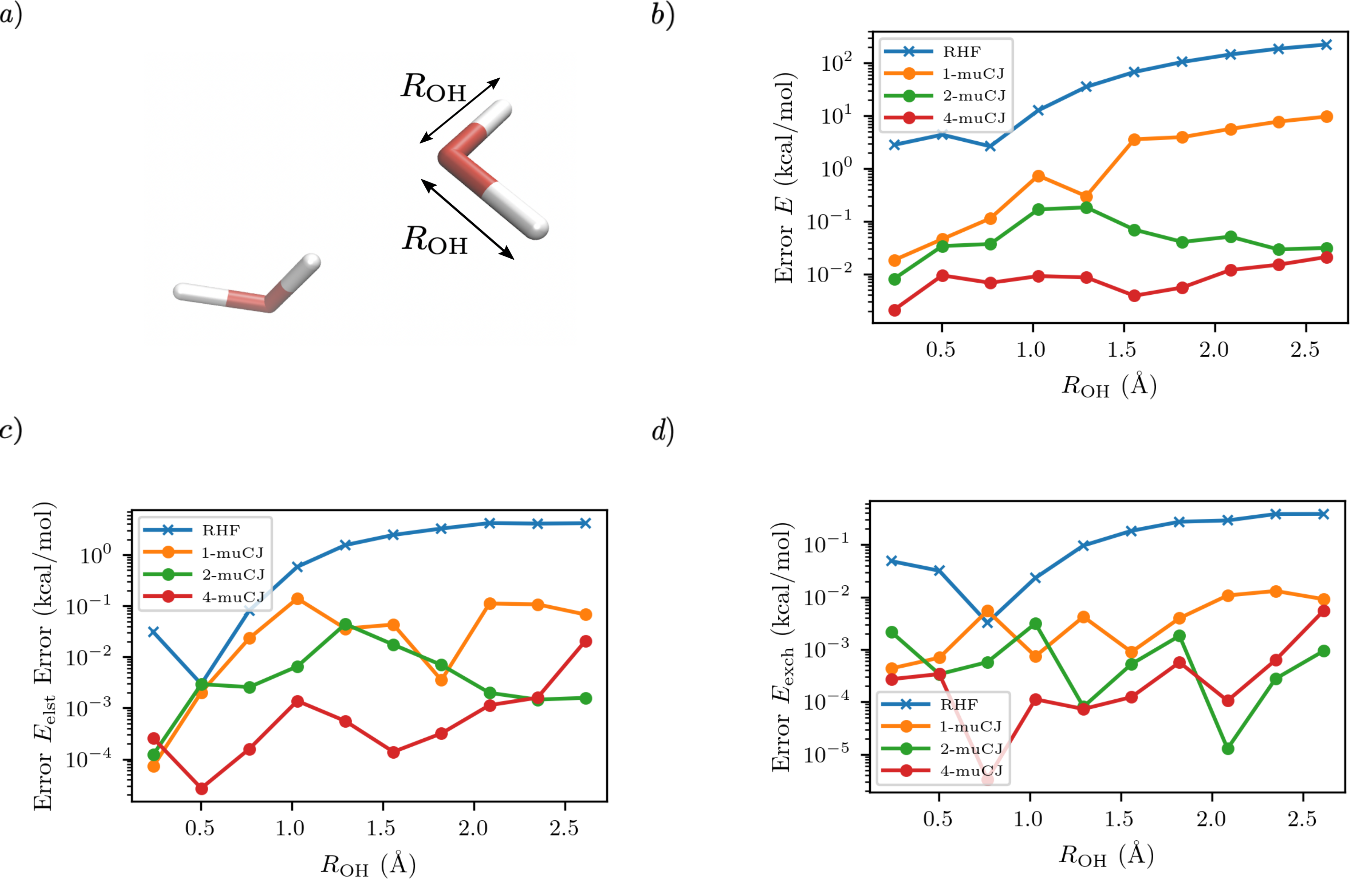}
    \caption{(a) Geometry of the water dimer studied with the OH bond length of single monomer distance labelled as $R_{\mathrm{OH}}$. The intermonomer separation is fixed throughout.
        (b) Absolute error in the stretched water monomer VQE total energy as a function of $R_{\mathrm{OH}}$ for different muCJ repetition factor $k$ in the $k$-muCJ ansatz. (c)-(d) Absolute errors relative to SAPT(CASCI) in the electrostatic and exchange energies calculated as a function of $R_{\mathrm{OH}}$ for different muCJ repetition factor $k$ in the $k$-muCJ ansatz.
    \label{fig:water}}
\end{figure*}

For the next test case we investigate the ability of SAPT(VQE) to address intermolecular interactions involving bond dissociation in one of the monomers.
We calculate the SAPT electrostatic and exchange energies for two interacting water molecules (displayed in \cref{fig:water} (a)), one in its equilibrium geometry and in the other we symmetrically stretch the two OH bonds towards dissociation.
This is again a challenging case for VQE to capture a double bond breaking.
We used the 6-31G basis set and chose a (6e, 6o) active space for water from HOMO-2 to LUMO+2.
Again, we see in \cref{fig:water} that RHF qualitatively fails to capture either the total energy of the stretched water monomer or the SAPT energy components of the dimer system.
On the other hand, while sizeable errors are present in the VQE total energy ($>$10 kcal/mol) in the stretched system for $k=1$, we find sub \mbox{0.1 kcal/mol} accuracy in the SAPT energy components.

It is interesting to note that while the quality of the total energy necessarily improves when the circuit repetition factor ($k$) is increased (assuming a local minima is not arrived at during optimization), some non-monotonic behaviour is observed in the individual SAPT energy components.
This is probably due to the difficulty in tightly converging the VQE total energy the log scale in these figures.

These two examples provide strong evidence that SAPT(VQE) yields
significantly lower absolute errors in the target interaction energy
contributions vs. those of the corresponding monomer total energies. This
helps to reduce the depth of circuits necessary to achieve accurate interaction energy components.

\subsection{Protein-Ligand Interactions}

\begin{figure*}
    \includegraphics[width=\textwidth]{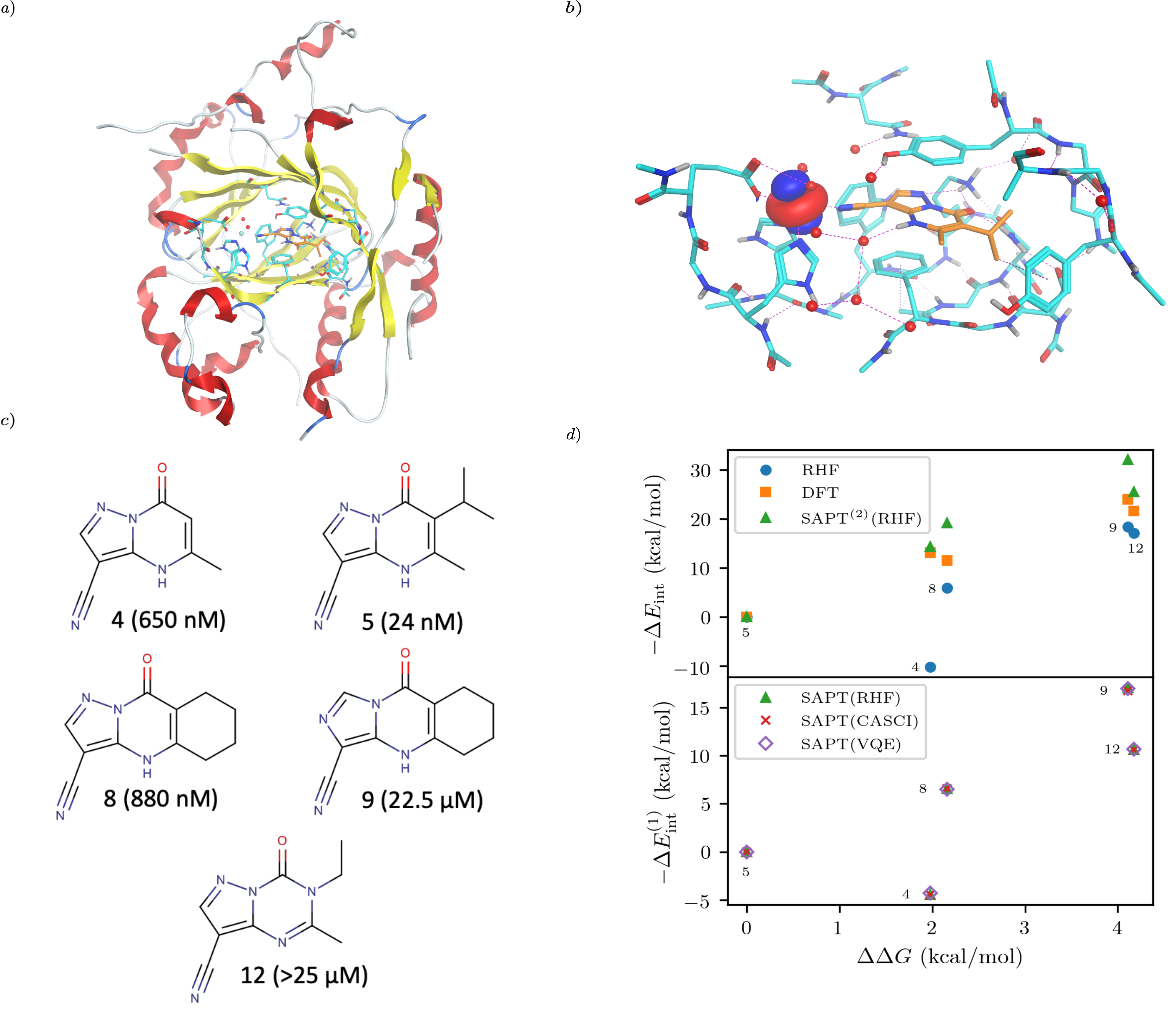}
    \caption{(a) KDM5A protein structure \citep{Labadie2016,Berman2000} (b) representative minimal active space molecular orbitals produced by the AVAS procedure, for the protein cutout described in the main text with ligand 5 colored as orange (c) 5 KDM5A inhibitors from Ref.~\citep{Gehling2016}, (d) Upper Panel: Interaction energy differences computed using the super molecular approach (DFT(B3YLP) and RHF) and SAPT$^{(2)}$(RHF) as a function of the difference in experimental free energies (see main text for definition). Lower Panel: First-order SAPT(VQE), SAPT(CASCI) and SAPT(RHF) first order interaction energy differences as a function of the differences in experimental free energies of binding.\label{fig:kdma5}}
\end{figure*}

For our final example we examine the ability of our implementation of SAPT(VQE) to tackle large protein-ligand interactions, specifically lysine-specific demethylase 5A (KDM5A) depicted in \cref{fig:kdma5} (a).
KDM5A is believed to be relevant for human cancers \citep{Yang2021} and contains a metal center [Fe(II)] which may pose a challenge for classical electronic structure theory methods.
To make the problem tractable, we used a model system of the binding site which was cut out from the full binding domain of KDM5A and focuses on the immediate surrounding of metal ion and ligand. The detailed preparation of the model system is described in the Supporting Information.
This final structures contained each 380 atoms and were studied with DFT using a 6-31G basis set and the B3LYP functional \citep{BeckeB3LYP1993} using the Gaussian software package \citep{g09}. 
The ligand and all atoms in radius of 4.5~\AA\ of the ligand were further relaxed with the oxygen atoms of the water and the iron atom been keeping fixed.
The structures have an electronic size of (1482e, 2214o).
The protein-ligand system is visualized in \cref{fig:kdma5} (b).
We computed the first order SAPT(VQE) energy contributions for 5 different inhibitors from Ref.~\cite{Gehling2016} and shown in \cref{fig:kdma5} (c) to assess how important multi-reference effects were for describing the interaction energy with this protein.
We treated each of the ligands at the RHF level and KDM5A using VQE.

To construct a potentially representative active space for KDM5A we used the AVAS procedure \citep{Sayfutyarova2017}, and first built a relatively large space by including the $3d$ orbitals from the iron center, the oxygen $2p$ orbitals from the nearest two water molecules and the $2p$ orbitals from the neighbouring oxygen atom associated with the glutamic acid, as well as two $2p$ orbitals from the neighbouring nitrogen atoms from two neighbouring histidines.
A representative Fe $3d$-like orbital from this procedure is depicted in \cref{fig:kdma5} (b).
This leads to an active space size from the AVAS procedure of (36e, 27o), which would require a 54 qubit quantum computer, and is thus outside the reach of current simulators and hardware.
To reduce the size of the active space we performed a loosely converged ($\varepsilon = 1\times10^{-4})$ selected heat bath configuration interaction computation (SHCI) \citep{Holmes2016,Sharma2017} in this active space.
We then constructed a smaller active space using the SHCI natural orbitals (NOs), only keeping those NOs with occupation $0.02 \le n_i \le 1.97$.
This leads to a smaller active space of (8e,8o), for the low-spin configuration, corresponding to a 16 qubits VQE simulation which is possible using simulators and within reach of current hardware.
The natural orbital occupation structure and dominant single reference nature of SHCI ground state suggests that the KDM5A problem will not be challenging for VQE.

In order to connect with experiment we score the ligands based on the difference of their interaction energy relative to a reference compound
\begin{equation}
    \Delta E_{\mathrm{int}}(x) = E_{\mathrm{int}}(x_\mathrm{ref}) - E_{\mathrm{int}}(x)
\end{equation}
where $x$ are the ligand labels from \cref{fig:kdma5} (c) and we choose $x_{\mathrm{ref}} = 5$ as the reference point as it is experimentally the most potent ligand from the subset of ligands taken from Ref.~\citenum{Gehling2016}.
We then compare this energy difference to differences in the experimental free energy of binding \citep{Yung-Chi1973}
\begin{equation}
    \Delta \Delta G = \Delta G(x_\mathrm{ref}) - \Delta G(x) = -RT \log \left(\frac{\mathrm{IC}_{50}(x_\mathrm{ref})}{\mathrm{IC}_{50}(x)}\right),
\end{equation}
where $R$ is the gas constant, the temperature $T$ is taken to be room temperature and we took the IC$_{50}$ values from Ref.~\citenum{Gehling2016}.
Note that we are comparing two different quantities, namely the theoretical interaction energy difference and the experimental free energy of binding and thus we cannot necessarily expect any quantitative relationship.
For example, the interaction energies account for one leg of the thermodynamic cycle while the free energy of binding accounts for the the whole cycle.
In addition, the number of water molecules surrounding the binding site may depend on the specific ligand being studied, with larger ligands replacing more water molecules.
While neither of these effects was considered here it is interesting to see what relationship if any these numbers have to one another.

The results of these computations are shown in \cref{fig:kdma5} (d).
To get some insight into the accuracy of existing approaches we first present supermolecular DFT(B3LYP), RHF and SAPT$^{(2)}$(RHF) the results of which are shown in the upper panel of \cref{fig:kdma5} (d).
Note that here SAPT$^{(2)}$(RHF) contains both first- and second-order SAPT energy contributions \citep{Parrish2018} and were performed in a dimer-centered basis set, this in contrast to the rest of the results in this paper which only considered the first-order SAPT contributions in a monomer-centered basis.
The DFT results are not counterpoise corrected but we found it unimportant for the present system.
We find that supermolecular RHF incorrectly predicts ligand-4 to have the largest interaction energy suggesting that a more accurate method is required.
This observation is confirmed by the SAPT(RHF) and supermolecular DFT results which correctly score ligand-5 as the most potent in this set.

In the lower panel of \cref{fig:kdma5} (d) we plot the first order interaction energy difference, $\Delta E_\mathrm{int}^{(1)}$ computed through SAPT(RHF), SAPT(CASCI) and SAPT(VQE) all computed using a monomer-centered basis.
The first thing we see is the excellent agreement between SAPT(CASCI) and SAPT(VQE) for each ligand considered.
Note that we used the 1-muCJ ansatz for the VQE computations which offers further evidence of the error reduction capabilities of SAPT.
Apart from this excellent agreement we also see that there is little difference between SAPT(RHF) and SAPT(CASCI).
We found that the maximum difference between SAPT(CASCI) and SAPT(RHF) was roughly 1-2 kcal/mol, however this error nearly cancels when looking at the $\Delta E_\mathrm{int}$ and is thus not visible on the scale of the plot.
Ultimately this is not surprising given that the problem was found to be not strongly correlated.

Finally, we see that first order SAPT is not accurate enough to correctly score the respective ligands with the missing induction and dispersion components proving critical to achieving this.
Indeed the first order interaction energy for compound 5 is positive and thus unbound at this level of theory.
In particular, from the full SAPT$^{(2)}$(RHF) energy decomposition (see Supplementary Information) we find that the induction components for compound 4 and compound 5 are similar in magnitude while the dispersion component for compound 5 is roughly 1.4 times that of compound 4.
This large dispersion energy ensures that ligand 5 has the larger (absolute) interaction energy.
Nevertheless, we see that SAPT(VQE) can potentially tackle industrially relevant drug design problems, particularly in cases where it is not known a-priori how strongly correlated the ground state is.

\section{Conclusion}

We described the theory and implementation of SAPT on a NISQ-era quantum computer, focusing on the efficient implementation and classical workflows necessary to tackle industrially relevant problems in drug design.
We derived in detail the equations necessary for an active space formulation of SAPT(VQE) that requires only the one- and two-particle reduced density matrices measured on a quantum computer. The SAPT(VQE) components are computed as an efficient classical post-processing step.
This classical post-processing step, written in terms of optimized quantum chemical primitives, was shown capable of simulating systems with hundreds of atoms and potentially 100s of basis functions in the active space.

Beyond an efficient classical implementation, we found (through ideal VQE simulations) that SAPT naturally reduces the error incurred by approximately solving the Schr{\"o}dinger equation on a quantum computer, which we attribute to the theory directly computing energy differences.
This fact coupled with a monomer basis formulation will substantially reduce the resource requirements for computing binding energies of large protein-ligand interactions on NISQ-era quantum computers.
In particular, sub-kcal/mol accuracy in the energy components for the electrostatic and exchange energies can be computed using coarse VQE wavefunctions, which otherwise exhibit gross errors ($>$ 10 kcal/mol) in the total energy.
Given the practical challenges associated with optimizing VQE wavefunctions on current hardware we believe SAPT may help to extend the scope of the method due to the apparent ability to use low depth VQE ansatzes like 1-uCJ.
SAPT(VQE) appears to offer, then, a reduction in the quantum resources required compared
to a simple supermolecular VQE for the computation of protein-ligand interaction energies. 
This reduction in qubit count and circuit depth naturally yields a lower
precision requirement on the quantum circuit and thus has the potential to enable the simulation of 
larger systems on NISQ hardware than what is possible with supermolecular VQE implementations.

In the future, it will be critical to determine how robust SAPT(VQE) is to noise channels either through modelling or real hardware experiments.
Another important question will be how to reduce the measurement overhead of accumulating the one- and two-particle reduced density matrices, a tentative solution to which is sketched in \cref{app:elst}.
A natural extension will be to determine the second-order SAPT terms that would allow for accurate interaction energies and induction and dispersion energy components to be computed.
Other interesting questions are how to increase the quantitative accuracy of the method by including correlation out of the active space \citep{Takeshita2020,Huggins2021} and to gather more challenging drug-protein systems potentially containing multiple metal centers.

\begin{acknowledgments}
QC Ware Corp.~acknowledges generous funding from Boehringer Ingelheim for the
undertaking of this project. We thank Clemens Utschig-Utschig and Christofer
Tautermann for insightful discussions.
\end{acknowledgments}

\section*{Conflict of Interest}
FDM, RMP, and ARW own stock/options in QC Ware Corp.

\appendix

\section{Full Active Space SAPT Exchange Expressions\label{app:active}}
In this appendix we give full expressions for the remaining exchange energy contributions ($T_3$ and $T_4$) not given in \cref{sec:active_space}.
For notational simplicity we will drop bars from $\gamma$ and $\Gamma$ under the understanding that all density matrices are spin-summed and remove the tildes from the generalized $J$ and $K$ matrices.

$T_3$ is very similar to $T_2$ and we have
\begin{equation}
    \begin{split}
        T_3 = &T_3[\gamma_B^c, \Gamma_A^{ac}] + T_3[\gamma_B^c, \Gamma_A^{cc}] + \\
              &T_3[\gamma_B^a, \Gamma_A^{ac}]+ T_3[\gamma_B^c, \Gamma_A^{aa}] + \\
              &T_3[\gamma_B^a, \Gamma_A^{aa}],
    \end{split}
\end{equation}
where
\begin{equation}
\begin{split}
&T_3[\gamma_B^c, \Gamma_A^{ac}]
=
\\
&4 S_{ji} C_{\nu j}
(J[\gamma^a_{\mu\mu'}]_{\nu\mu''}
-\frac{1}{2}K[\gamma^a_{\mu\mu'}]_{\nu\mu''}) C_{\mu'' i}\\
+
&2 \gamma_{tt'} S_{jt} C_{\nu j}
(J[\gamma^c_{\mu\mu'}]_{\nu\mu''}
-
\frac{1}{2}K[\gamma^c_{\mu\mu'}]_{\nu\mu''})
C_{\mu'' t'},
\end{split}
\end{equation}
\begin{equation}
    T_3[\gamma_B^c,\Gamma_A^{cc}] = 2\gamma_{uu'} S_{ui'} C_{\nu u}\left(J[\gamma^c_{\mu\mu'}]-\frac{1}{2}K[\gamma^{c}_{\mu\mu'}]\right)C_{\mu''i},
\end{equation}
\begin{equation}
\begin{split}
&T_3[\gamma_B^a, \Gamma_A^{ac}]
=
\\
&2 \gamma_{uu'} S_{u'i} C_{\nu u}
(J[\gamma^a_{\mu\mu'}]_{\nu\mu''}
-\frac{1}{2}K[\gamma^a_{\mu\mu'}]_{\nu\mu''}) C_{\mu'' i}\\
+
& \gamma_{uu'} \gamma_{tt'} S_{u't} C_{\nu u}
(J[\gamma^c_{\mu\mu'}]_{\nu\mu''}
-
\frac{1}{2}K[\gamma^c_{\mu\mu'}]_{\nu\mu''})
C_{\mu'' t'},
\end{split}
\end{equation}
\begin{equation}
    T_3[\gamma_B^c, \Gamma_A^{aa}] = 2 \Gamma^{tt'}_{t''t'''} \tilde{v}_{tt'}^{t'' t'''},
\end{equation}
and, lastly,
\begin{equation}
T_3[\gamma_B^a, \Gamma_A^{aa}]
=
\gamma_{uu'}
\Gamma^{tt'}_{t''t'''} S_{u't'} \tilde{v}_{tu}^{t''t'''}.
\end{equation}

For $T_4$ there are sixteen terms in total.

First we have
\begin{equation}
\begin{split}
&T_4[\Gamma^{aa}_A,\Gamma^{aa}_B] = \\
&\Gamma_{tt''t't'''}\Gamma_{uu''u'u'''}S_{t'''u'}S_{t'u'''}\tilde{v}_{tt''uu''},
\end{split}
\end{equation}
which can't be simplified further.
Next we have
\begin{equation}
\begin{split}
&T_4[\Gamma^{aa}_A,\Gamma^{ac}_B] = \\
&\Gamma_{tt''t't'''}\gamma_{uu''}\gamma_{jj''}S_{t'''j}S_{t'j''}\tilde{v}_{tt''uu''}\\
&-\frac{1}{2}\Gamma_{tt''t't'''}\gamma_{uu''}\gamma_{jj''}S_{t'''j}S_{t'u''}\tilde{v}_{tt''uj''}\\
&+\Gamma_{tt''t't'''}\gamma_{uu''}\gamma_{jj''}S_{t'''u}S_{t'u''}\tilde{v}_{tt''jj''}\\
&-\frac{1}{2}\Gamma_{tt''t't'''}\gamma_{uu''}\gamma_{jj''}S_{t'''u}S_{t'j''}\tilde{v}_{tt''ju''},
\end{split}
\end{equation}
which can be written as
\begin{equation}
\begin{split}
&T_4[\Gamma^{aa}_A,\Gamma^{ac}_B] = \\
&2\Gamma_{tt''t't'''}S_{t'''j}S_{t'j}J[\gamma_{uu''}]_{tt''}\\
&-\Gamma_{tt''t't'''}\gamma_{uu''}S_{t'u''}\tilde{v}_{tt''ut'''}\\
&+\Gamma_{tt''t't'''}\gamma_{uu''}S_{t'''u}S_{t'u''}J[\gamma_{jj''}]_{tt''}\\
&-\Gamma_{tt''t't'''}\gamma_{uu''}S_{t'''u}\tilde{v}_{tt''t'u''}.
\end{split}
\end{equation}
Similarly
\begin{equation}
\begin{split}
&T_4[\Gamma^{ac}_A,\Gamma^{aa}_B] = \\
&\gamma_{tt''}\gamma_{ii''}\Gamma_{uu''u'u'''}S_{i''u'}S_{iu'''}\tilde{v}_{tt''uu''}\\
&-\frac{1}{2}\gamma_{tt''}\gamma_{ii''}\Gamma_{uu''u'u'''}S_{t''u'}S_{iu'''}\tilde{v}_{ti''uu''}\\
&+\gamma_{tt''}\gamma_{ii''}\Gamma_{uu''u'u'''}S_{t''u'}S_{tu'''}\tilde{v}_{ii''uu''}\\
&-\frac{1}{2}\gamma_{tt''}\gamma_{ii''}\Gamma_{uu''u'u'''}S_{i''u'}S_{tu'''}\tilde{v}_{it''uu''},
\end{split}
\end{equation}
which can be written as
\begin{equation}
\begin{split}
&T_4[\Gamma^{ac}_A,\Gamma^{aa}_B] = \\
&2\Gamma_{uu''u'u'''}S_{iu'}S_{iu'''}J[\gamma_{tt''}]_{uu''}\\
&-\gamma_{tt''}\Gamma_{uu''u'u'''}S_{t''u'}\tilde{v}_{tu'''uu''}\\
&+\gamma_{tt''}\Gamma_{uu''u'u'''}S_{t''u'}S_{tu'''}J[\gamma_{ii''}]_{uu''}\\
&-\gamma_{tt''}\Gamma_{uu''u'u'''}S_{tu'''}\tilde{v}_{u't''uu''}.
\end{split}
\end{equation}

Next
\begin{equation}
\begin{split}
&T_4[\Gamma^{aa}_A,\Gamma^{cc}_B] = \\
&\Gamma_{tt''t't'''}\gamma_{jj''}\gamma_{j'j'''}S_{t'''j'}S_{t'j'''}\tilde{v}_{tt''jj''}\\
&-\frac{1}{2}\Gamma_{tt''t't'''}\gamma_{jj'''}\gamma_{j'j''}S_{t'''j'}S_{t'j'''}\tilde{v}_{tt''jj''},
\end{split}
\end{equation}
which can be written as
\begin{equation}
\begin{split}
&T_4[\Gamma^{aa}_A,\Gamma^{cc}_B] = \\
&2\Gamma_{tt''t't'''}S_{t'''j'}S_{t'j'}J[\gamma_{jj''}]_{tt''}\\
&-
\Gamma_{tt''t't'''}\tilde{v}_{tt''t't'''},\\
\end{split}
\end{equation}
and similarly
\begin{equation}
\begin{split}
&T_4[\Gamma^{cc}_A,\Gamma^{aa}_B] = \\
&\gamma_{ii''}\gamma_{i'i'''}\Gamma_{uu''u'u'''}S_{i'''u'}S_{i'u'''}\tilde{v}_{ii''uu''}\\
&-\frac{1}{2}\gamma_{ii'''}\gamma_{i'i''}\Gamma_{uu''u'u'''}S_{i'''u'}S_{i'u'''}\tilde{v}_{ii''uu''},
\end{split}
\end{equation}
which can be written as
\begin{equation}
\begin{split}
&T_4[\Gamma^{cc}_A,\Gamma^{aa}_B] = \\
&2\Gamma_{uu''u'u'''}S_{i'u'}S_{i'u'''}J[\gamma_{ii''}]_{uu''}\\
&-2\Gamma_{uu''u'u'''}\tilde{v}_{u'u'''uu''}.
\end{split}
\end{equation}

For the AC-AC contribution we get sixteen terms
\begin{equation}
\begin{split}
&T_4[\Gamma^{ac}_A,\Gamma^{ac}_B] = \\
&\gamma_{tt''}\gamma_{ii''}\gamma_{uu''}\gamma_{jj''}S_{i''j}S_{ij''}\tilde{v}_{tt''uu''}\\
&-\frac{1}{2}\gamma_{tt''}\gamma_{ii''}\gamma_{uu''}\gamma_{jj''}S_{i''j}S_{iu''}\tilde{v}_{tt''uj''}\\
&+\gamma_{tt''}\gamma_{ii''}\gamma_{uu''}\gamma_{jj''}S_{i''u}S_{iu''}\tilde{v}_{tt''jj''}\\
&-\frac{1}{2}\gamma_{tt''}\gamma_{ii''}\gamma_{uu''}\gamma_{jj''}S_{i''u}S_{ij''}\tilde{v}_{tt''ju''}\\
&-\frac{1}{2}\gamma_{tt''}\gamma_{ii''}\gamma_{uu''}\gamma_{jj''}S_{t''j}S_{ij''}\tilde{v}_{ti''uu''}\\
&+\frac{1}{4}\gamma_{tt''}\gamma_{ii''}\gamma_{uu''}\gamma_{jj''}S_{t''j}S_{iu''}\tilde{v}_{ti''uj''}\\
&-\frac{1}{2}\gamma_{tt''}\gamma_{ii''}\gamma_{uu''}\gamma_{jj''}S_{t''u}S_{iu''}\tilde{v}_{ti''jj''}\\
&+\frac{1}{4}\gamma_{tt''}\gamma_{ii''}\gamma_{uu''}\gamma_{jj''}S_{t''u}S_{ij''}\tilde{v}_{ti''ju''}\\
&+\gamma_{tt''}\gamma_{ii''}\gamma_{uu''}\gamma_{jj''}S_{t''j}S_{tj''}\tilde{v}_{ii''uu''}\\
&-\frac{1}{2}\gamma_{tt''}\gamma_{ii''}\gamma_{uu''}\gamma_{jj''}S_{t''j}S_{tu''}\tilde{v}_{ii''uj''}\\
&+\gamma_{tt''}\gamma_{ii''}\gamma_{uu''}\gamma_{jj''}S_{t''u}S_{tu''}\tilde{v}_{ii''jj''}\\
&-\frac{1}{2}\gamma_{tt''}\gamma_{ii''}\gamma_{uu''}\gamma_{jj''}S_{t''u}S_{tj''}\tilde{v}_{ii''ju''}\\
&-\frac{1}{2}\gamma_{tt''}\gamma_{ii''}\gamma_{uu''}\gamma_{jj''}S_{i''j}S_{tj''}\tilde{v}_{it''uu''}\\
&+\frac{1}{4}\gamma_{tt''}\gamma_{ii''}\gamma_{uu''}\gamma_{jj''}S_{i''j}S_{tu''}\tilde{v}_{it''uj''}\\
&-\frac{1}{2}\gamma_{tt''}\gamma_{ii''}\gamma_{uu''}\gamma_{jj''}S_{i''u}S_{tu''}\tilde{v}_{it''jj''}\\
&+\frac{1}{4}\gamma_{tt''}\gamma_{ii''}\gamma_{uu''}\gamma_{jj''}S_{i''u}S_{tj''}\tilde{v}_{it''ju''},
\end{split}
\end{equation}
which can be simplified to
\begin{equation}
\begin{split}
&T_4[\Gamma^{ac}_A,\Gamma^{ac}_B] = \\
&4\gamma_{tt''}S_{ij}S_{ij}J[\gamma_{uu''}]_{tt''}\\
&-2\gamma_{uu''}S_{ij}S_{iu''}J[\gamma_{tt''}]_{uj}\\
&+2\gamma_{tt''}\gamma_{uu''}S_{iu}S_{iu''}J[\gamma_{jj''}]_{tt''}\\
&-2\gamma_{uu''}S_{iu}S_{ij}J[\gamma_{tt''}]_{ju''}\\
&-2\gamma_{tt''}S_{t''j}S_{ij}J[\gamma_{uu''}]_{ti}\\
&+\gamma_{uu''}S_{iu''}K[\gamma_{tt''}S_{t''j}]_{iu}\\
&-\gamma_{tt''}\gamma_{uu''}S_{t''u}S_{iu''}J[\gamma_{jj''}]_{ti}\\
&+S_{ij}K[\gamma_{tt''}S_{t''u}\gamma_{uu''}]_{ij}\\
&+4\gamma_{tt''}S_{t''j}S_{tj}J[\gamma_{uu''}]_{ii}\\
&-\gamma_{tt''}\gamma_{uu''}S_{t''j}S_{tu''}J[\gamma_{ii''}]_{uj}\\
&+2\gamma_{tt''}\gamma_{uu''}S_{t''u}S_{tu''}J[\gamma_{jj''}]_{ii}\\
&-\gamma_{tt''}\gamma_{uu''}S_{t''u}S_{tj}J[\gamma_{ii''}]_{ju''}\\
&-2\gamma_{tt''}S_{ij}S_{tj}J[\gamma_{uu''}]_{it''}\\
&+\gamma_{tt''}\gamma_{uu''}S_{tu''}K[S_{ij}]_{t''u}\\
&-\gamma_{tt''}\gamma_{uu''}S_{iu}S_{tu''}J[\gamma_{jj''}]_{it''}\\
&+
\gamma_{tt''}S_{tj}K[\gamma_{uu''}S_{iu}]_{t''j}.
\end{split}
\end{equation}

Next we have
\begin{equation}
\begin{split}
&T_4[\Gamma^{ac}_A,\Gamma^{cc}_B] = \\
&\gamma_{tt''}\gamma_{ii''}\gamma_{jj''}\gamma_{j'j'''}S_{i''j'}S_{ij'''}\tilde{v}_{tt''jj''}\\
&-\frac{1}{2}\gamma_{tt''}\gamma_{ii''}\gamma_{jj'''}\gamma_{j'j''}S_{i''j'}S_{ij'''}\tilde{v}_{tt''jj''}\\
&-\frac{1}{2}\gamma_{tt''}\gamma_{ii''}\gamma_{jj''}\gamma_{j'j'''}S_{t''j'}S_{ij'''}\tilde{v}_{ti''jj''}\\
&+\frac{1}{4}\gamma_{tt''}\gamma_{ii''}\gamma_{jj'''}\gamma_{j'j''}S_{t''j'}S_{ij'''}\tilde{v}_{ti''jj''}\\
&+\gamma_{tt''}\gamma_{ii''}\gamma_{jj''}\gamma_{j'j'''}S_{t''j'}S_{tj'''}\tilde{v}_{ii''jj''}\\
&-\frac{1}{2}\gamma_{tt''}\gamma_{ii''}\gamma_{jj'''}\gamma_{j'j''}S_{t''j'}S_{tj'''}\tilde{v}_{ii''jj''}\\
&-\frac{1}{2}\gamma_{tt''}\gamma_{ii''}\gamma_{jj''}\gamma_{j'j'''}S_{i''j'}S_{tj'''}\tilde{v}_{it''jj''}\\
&+\frac{1}{4}\gamma_{tt''}\gamma_{ii''}\gamma_{jj'''}\gamma_{j'j''}S_{i''j'}S_{tj'''}\tilde{v}_{it''jj''},
\end{split}
\end{equation}
which can be written as
\begin{equation}
\begin{split}
&T_4[\Gamma^{ac}_A,\Gamma^{cc}_B] = \\
&8 S_{ij'}S_{ij'}J[\gamma_{tt''}]_{jj}\\
&-4 S_{ij'}S_{ij}J[\gamma_{tt''}]_{jj'}\\
&-2 \gamma_{tt''}S_{t''j'}S_{ij'}J[\gamma_{jj''}]_{ti}\\
&+2 S_{ij}K[\gamma_{tt''}S_{t''j'}]_{ij}\\
&+4\gamma_{tt''}S_{t''j'}S_{tj'}J[\gamma_{jj''}]_{ii}\\
&-2\gamma_{tt''}S_{t''j'}S_{tj}J[\gamma_{ii''}]_{jj'}\\
&-2\gamma_{tt''}S_{ij'}S_{tj'}J[\gamma_{jj''}]_{it''}\\
&+2\gamma_{tt''}S_{tj}K[S_{ij'}]_{t''j},
\end{split}
\end{equation}
and similarly
\begin{equation}
\begin{split}
&T_4[\Gamma^{cc}_A,\Gamma^{ac}_B] = \\
&\gamma_{ii''}\gamma_{i'i'''}\gamma_{uu''}\gamma_{jj''}S_{i'''j}S_{i'j''}\tilde{v}_{ii''uu''}\\
&-\frac{1}{2}\gamma_{ii''}\gamma_{i'i'''}\gamma_{uu''}\gamma_{jj''}S_{i'''j}S_{i'u''}\tilde{v}_{ii''uj''}\\
&+\gamma_{ii''}\gamma_{i'i'''}\gamma_{uu''}\gamma_{jj''}S_{i'''u}S_{i'u''}\tilde{v}_{ii''jj''}\\
&-\frac{1}{2}\gamma_{ii''}\gamma_{i'i'''}\gamma_{uu''}\gamma_{jj''}S_{i'''u}S_{i'j''}\tilde{v}_{ii''ju''}\\
&-\frac{1}{2}\gamma_{ii'''}\gamma_{i'i''}\gamma_{uu''}\gamma_{jj''}S_{i'''j}S_{i'j''}\tilde{v}_{ii''uu''}\\
&+\frac{1}{4}\gamma_{ii'''}\gamma_{i'i''}\gamma_{uu''}\gamma_{jj''}S_{i'''j}S_{i'u''}\tilde{v}_{ii''uj''}\\
&-\frac{1}{2}\gamma_{ii'''}\gamma_{i'i''}\gamma_{uu''}\gamma_{jj''}S_{i'''u}S_{i'u''}\tilde{v}_{ii''jj''}\\
&+\frac{1}{4}\gamma_{ii'''}\gamma_{i'i''}\gamma_{uu''}\gamma_{jj''}S_{i'''u}S_{i'j''}\tilde{v}_{ii''ju''}\\
\end{split}
\end{equation}
\begin{equation}
\begin{split}
&T_4[\Gamma^{cc}_A,\Gamma^{ac}_B] = \\
&8 S_{i'j}S_{i'j}J[\gamma_{uu''}]_{ii}\\
&-
2\gamma_{uu''}S_{i'j}S_{i'u''}J[\gamma_{ii''}]_{uj}\\
&+4\gamma_{uu''}S_{i'u}S_{i'u''}J[\gamma_{jj''}]_{ii}\\
&-2\gamma_{uu''}S_{i'u}S_{i'j}J[\gamma_{ii''}]_{ju''}\\
&-4S_{ij}S_{i'j}J[\gamma_{uu''}]_{ii'}\\
&+2\gamma_{uu''}S_{i'u''}K[S_{ij}]_{i'u}\\
&-2\gamma_{uu''}S_{iu}S_{i'u''}J[\gamma_{jj''}]_{ii'}\\
&+2S_{i'j}K[\gamma_{uu''}S_{iu}]_{i'j}.
\end{split}
\end{equation}
Finally, we have
\begin{equation}
\begin{split}
&T_4[\Gamma^{cc}_A,\Gamma^{cc}_B] = \\
&\gamma_{ii''}\gamma_{i'i'''}\gamma_{jj''}\gamma_{j'j'''}S_{i'''j'}S_{i'j'''}\tilde{v}_{ii''jj''}\\
&-\frac{1}{2}\gamma_{ii''}\gamma_{i'i'''}\gamma_{jj'''}\gamma_{j'j''}S_{i'''j'}S_{i'j'''}\tilde{v}_{ii''jj''}\\
&-\frac{1}{2}\gamma_{ii'''}\gamma_{i'i''}\gamma_{jj''}\gamma_{j'j'''}S_{i'''j'}S_{i'j'''}\tilde{v}_{ii''jj''}\\
&+\frac{1}{4}\gamma_{ii'''}\gamma_{i'i''}\gamma_{jj'''}\gamma_{j'j''}S_{i'''j'}S_{i'j'''}\tilde{v}_{ii''jj''},
\end{split}
\end{equation}
which can be written in terms of $J$ and $K$ matrices as was done for the SAPT(HF).

\section{Quantum-Classical Optimization for Electrostatic Energy \label{app:elst}}

When only monomer $A$ is defined to be quantum, an interesting alternative approach exists to evaluating the one-particle density matrix on the quantum computer.
We may first classically form the contributions from the nuclei of monomer $A$,
\begin{equation}
E_{\mathrm{elst,u}}^{\mathrm{VQE}}
\leftarrow
\sum_{A, B}
\frac{Z_{A} Z_{B}}{r_{AB}}
+
\sum_{A\nu \nu'}
(A | \nu \nu' )
\gamma_{\nu \nu'}
\end{equation}
We then classically form the image of the electrostatic potential of monomer $B$
in the atomic orbital basis of monomer $A$ as,
\begin{equation}
W_{\mu \mu'}
\equiv
\sum_{B}
(\mu \mu' | B)
+
J_{\mu \mu'} [\gamma_{\nu \nu'}]
\end{equation}
We can form the monomer $A$ core $\leftrightarrow$ monomer $B$ contributions
classically as,
\begin{equation}
E_{\mathrm{elst,u}}^{\mathrm{VQE}}
\leftarrow
\sum_{\mu \mu'}
\gamma_{\mu \mu'}^{\mathrm{Core}}
W_{\mu \mu'}
\end{equation}
We can then classically form the image of the electrostatic potential of monomer
$B$ in the active space molecular orbital basis of monomer $A$ as,
\begin{equation}
W_{tt'}
\equiv
\sum_{\mu \mu'}
C_{\mu t}
W_{\mu \mu'}
C_{\mu' t'}
\end{equation}
We can then diagonalize the $W_{tt'}$ operator to form,
\begin{equation}
W_{tt'}
=
\sum_{s}
U_{ts}
w_{s}
U_{t's}
\end{equation}
where $U_{ts}$ is $\mathcal{SO}(N_{t})$. In this ``electrostatic potential
natural orbital basis'' the remaining monomer $A$ active $\leftrightarrow$
monomer $B$ contributions can be evaluated by a single commuting group of
simultaneous $Z$-basis measurements,
\begin{align}
E_{\mathrm{elst,u}}^{\mathrm{VQE}}
&\leftarrow
\sum_{\mu \mu'}
\gamma_{\mu \mu'}^{a}
W_{\mu \mu'}
=
\sum_{t t'}
\gamma_{t t'}^{a}
W_{t t'}\\
&=
\sum_{tt'}
\langle \Omega_{A} |
t^\dagger 
t'
+
\bar t^\dagger 
\bar t'
| \Omega_{A} \rangle
W_{tt'}\\
&=
\sum_{s}
\langle \Omega_{A} |
\hat V^{\dagger} (U_{ts})
\left [
s^\dagger 
s
+
\bar s^\dagger 
\bar s
\right ]
\hat V (U_{ts})
| \Omega_{A} \rangle
w_{s}\\
&=
\sum_{s} w_{s}
-
\frac{1}{2}
\sum_{s}
\langle \Omega_{A} |
\hat V^{\dagger} (U_{ts})
\left [
Z_{s}
+
Z_{\bar s}
\right ]
\hat V (U_{ts})
| \Omega_{A} \rangle
w_{s}
\end{align}
This is in contrast to a naive implementation which would require $\mathcal{O}(N^2)$ circuit evaluations or an optimal method of $N/2$ circuits evaluations \citep{AIQuantum2020}.
Similar ideas can be explored for the exchange expression. Given that its structure mirrors that of a total energy evaluation (a 4-index tensor contracted with the two-particle reduced density matrix) it seems possible that a double factorization \citep{MottaDF2021,HugginsDF2021} approach may be possible.
These ideas will be explored at a later date.

\bibliography{refs.bib}


\end{document}


\definecolor{brickred}{rgb}{.72,0,0} 

\title{
Supplementary Information for Towards the Simulating Large Scale Protein-Ligand Interactions on NISQ-era Quantum Computers
}

\author{Fionn D.~Malone~\orcidlink{0000-0001-9239-0162}}
\author{Robert M.~Parrish~\orcidlink{0000-0002-2406-4741}}
\email{rob.parrish@qcware.com}
\author{Alicia R.~Welden~\orcidlink{0000-0002-2238-9825}}
\affiliation{QC Ware Corporation, Palo Alto, CA, 94301, USA}
\author{Thomas Fox~\orcidlink{0000-0002-1054-4701}}
\affiliation{Medicinal Chemistry, Boehringer Ingelheim Pharma GmbH \& Co. KG, Birkendorfer Stra{\ss}e 65, 88397 Biberach an der Ri\ss, Germany}
\author{Matthias Degroote~\orcidlink{0000-0002-8850-7708}}
\author{Elica ~Kyoseva~\orcidlink{0000-0002-9154-0293}}
\author{Nikolaj Moll~\orcidlink{0000-0001-5645-4667}}
\email{nikolaj.moll@boehringer-ingelheim.com}
\author{Raffaele Santagati~\orcidlink{0000-0001-9645-0580}}
\author{Michael Streif~\orcidlink{0000-0002-7509-4748}}
\affiliation{Quantum Lab, Boehringer Ingelheim, 55218 Ingelheim am Rhein, Germany}

\maketitle
\section{Data for Benzene--p-Benzyne dimer}

\begin{figure}[h!]
    \includegraphics{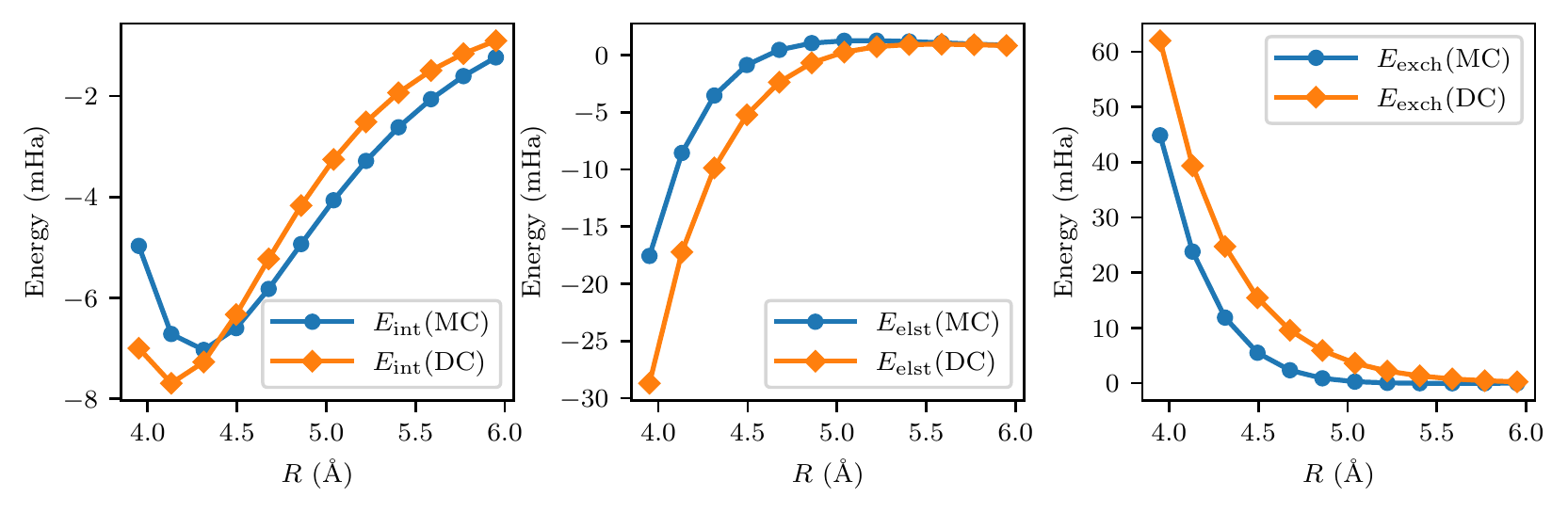}
    \caption{Comparison between monomer-centered (MC) and dimer-centered(DC) SAPT(RHF) energy components for the benzene--p-benzyne dimer. Note unlike other calculations presented in this work, these results were computed using the density-fitting version of SAPT \citep{HohensteinDFSAPT02010,HohensteinDFSAPT12011} implemented in the psi4 quantum chemistry package \citep{TurneyPSI42012}. \label{fig:pbp_breakdown}}
\end{figure}

In \cref{tab:pbp_elst} and \cref{tab:pbp_exch} we present results for the benzene--p-benzyne dimer considered in the main text. All calculations were performed in the (monomer-centered) cc-pVDZ basis set and a gradient threshold of $1\times10^{-6}$ for the L-BFGS-B solver in scipy.
For comparison in \cref{fig:pbp_breakdown} we compare the SAPT(RHF) energy components in the monomer-centered and dimer-centered basis sets. We found that in the monomer-centered basis set the exchange energy became a small negative number which is likely a numerical artifact but was consistent across different codes.

\begin{table}[h!]
\begin{tabular}{rrrrr}
    \hline
    $R$ (\AA) &       SAPT(RHF) &     SAPT(CASCI) &     SAPT(1-uCJ) &     SAPT(5-uCJ) \\
    \hline
    3.950000 & -0.017566 & -0.018136 & -0.018212 & -0.018162 \\
    4.131818 & -0.008560 & -0.009209 & -0.009319 & -0.009251 \\
    4.313636 & -0.003534 & -0.004184 & -0.004303 & -0.004138 \\
    4.495455 & -0.000865 & -0.001477 & -0.001594 & -0.001448 \\
    4.677273 &  0.000461 & -0.000097 & -0.000205 & -0.000129 \\
    4.859091 &  0.001052 &  0.000552 &  0.000454 &  0.000575 \\
    5.040909 &  0.001255 &  0.000813 &  0.000726 &  0.000796 \\
    5.222727 &  0.001267 &  0.000877 &  0.000800 &  0.000895 \\
    5.404545 &  0.001191 &  0.000848 &  0.000780 &  0.000835 \\
    5.586364 &  0.001082 &  0.000779 &  0.000720 &  0.000768 \\
    5.768182 &  0.000966 &  0.000699 &  0.000647 &  0.000683 \\
    5.950000 &  0.000855 &  0.000620 &  0.000573 &  0.000636 \\
    \hline
\end{tabular}
\caption{First order $E_{\mathrm{elst}}$ for benzene--p-benzyne dimer as a function of the center-to-center distance $R$ (in angstrom) for different levels of SAPT. Energies are in Hartree.\label{tab:pbp_elst}}
\end{table}

\begin{table}[h!]
\begin{tabular}{rrrrr}
    \hline
         $R$ &  SAPT(RHF) &  SAPT(CASCI) &  SAPT(1-uCJ) &  SAPT(5-uCJ) \\
         \hline
         3.950000 &   0.044729 &     0.046125 &     0.046124 &     0.046149 \\
         4.131818 &   0.023684 &     0.024907 &     0.024942 &     0.025003 \\
         4.313636 &   0.011746 &     0.012730 &     0.012727 &     0.012718 \\
         4.495455 &   0.005388 &     0.006068 &     0.006079 &     0.006039 \\
         4.677273 &   0.002230 &     0.002668 &     0.002682 &     0.002674 \\
         4.859091 &   0.000787 &     0.001053 &     0.001064 &     0.001045 \\
         5.040909 &   0.000196 &     0.000345 &     0.000357 &     0.000347 \\
         5.222727 &  -0.000008 &     0.000074 &     0.000080 &     0.000074 \\
         5.404545 &  -0.000055 &    -0.000012 &    -0.000009 &    -0.000012 \\
         5.586364 &  -0.000050 &    -0.000028 &    -0.000027 &    -0.000028 \\
         5.768182 &  -0.000034 &    -0.000023 &    -0.000022 &    -0.000023 \\
         5.950000 &  -0.000020 &    -0.000015 &    -0.000014 &    -0.000015 \\
         \hline
     \end{tabular}
\caption{First order $E_{\mathrm{exch}}$ for benzene--p-benzyne dimer as a function of the center-to-center distance $R$ (in angstrom) for different levels of SAPT. Energies are in Hartree.\label{tab:pbp_exch}}
\end{table}

\begin{figure}
    \centering
\begin{BVerbatim}
    C    -1.3940633    0.0000000   -2.0000000
    C    -0.6970468    1.2072378   -2.0000000
    C     0.6970468    1.2072378   -2.0000000
    C     1.3940633    0.0000000   -2.0000000
    C     0.6970468   -1.2072378   -2.0000000
    C    -0.6970468   -1.2072378   -2.0000000
    H    -2.4753995    0.0000000   -2.0000000
    H    -1.2382321    2.1435655   -2.0000000
    H     1.2382321    2.1435655   -2.0000000
    H     2.4753995    0.0000000   -2.0000000
    H     1.2382321   -2.1435655   -2.0000000
    H    -1.2382321   -2.1435655   -2.0000000
    --
    C     0.0000000    0.0000000    1.0590353
    C     0.0000000   -1.2060084    1.7576742
    C     0.0000000   -1.2071767    3.1515905
    C     0.0000000    0.0000000    3.8485751
    C     0.0000000    1.2071767    3.1515905
    C     0.0000000    1.2060084    1.7576742
    H     0.0000000   -2.1416387    1.2144217
    H     0.0000000   -2.1435657    3.6929953
    H     0.0000000    2.1435657    3.6929953
    H     0.0000000    2.1416387    1.2144217
\end{BVerbatim}
\caption{Coordinates for benzene and p-benzyne dimer given at a center-to-center distance of $R$ = 4.45 \AA.}
\end{figure}

\section{Data for H$_2$O}

\begin{figure}[h!]
    \includegraphics[scale=1.4]{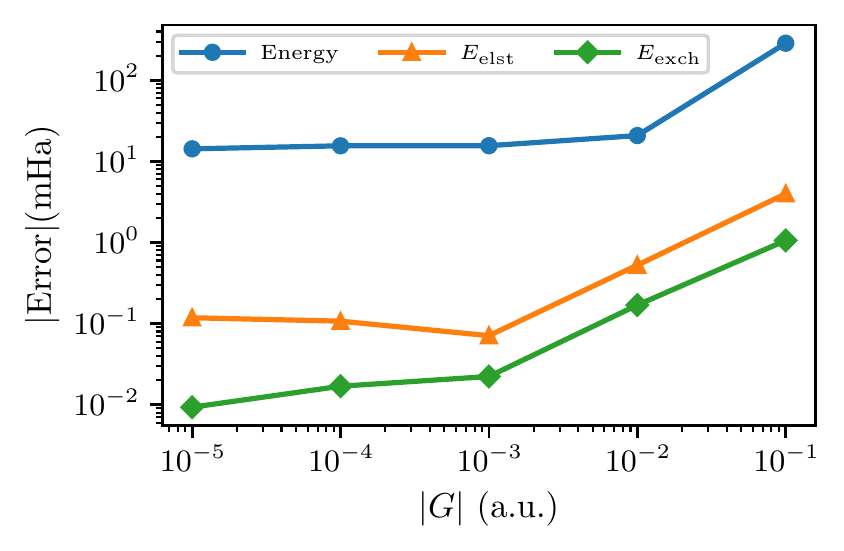}
    \caption{Convergence of errors (relative to CASCI) for the total energy, electrostatic and exchange energies when using the modified 1-uCJ ansatz as a function of the gradient threshold for the L-BFGS-B algorithm implemented in scipy (gtol). To obtain the data we first initialised the parameters randomly and fed the output parameters from one BFGS optimization at a larger threshold into the next optimization step with a smaller threshold. \label{fig:gtol_water}}
\end{figure}

In \cref{tab:rhf_water,tab:casic_water,tab:vqe_water} we present RHF, CASCI and VQE SAPT results for the water dimer considered as a function of the stretched OH bond length in one of the monomers.
We also provide for reference the total energy of the isolated water molecule undergoing the symmetric stretch.
All calculations were performed in the (monomer-centered) 6-31g basis set and a gradient threshold of $1\times10^{-6}$ for the L-BFGS-B solver in scipy or with a maximum iteration of 1000 whichever occurred first. In cases when the maximum iteration was reached first we found the norm of the gradient to be $\approx 1\times10^{-5}$.
In order to obtain a relatively smooth potential energy curve we started from the stretched geometry with randomly chosen parameters.
We then used these optimized parameters for the next more compressed geometry and so on.
To check the dependence of the energies on the quality of the VQE solution we plot in \cref{fig:gtol_water} the VQE total energy, electrostatic and exchange energies as a function of the gradient threshold. We see that the errors are quite well converged around $1\times10^{-3}$ and the SAPT components are quite insensitive to how well optimized the wavefunction is.
Finally in \cref{xyz:water} we provide the xyz coordinates for the water dimer at $R_{\mathrm{OH}} = 0.2397 $ \AA.

\begin{table}[h!]
\begin{tabular}{rrrr}
    \hline
    $R$ &     $E_{\mathrm{elst}}$ &    $E_{\mathrm{exch}}$ &        $E_{\mathrm{RHF}}$ \\
    \hline
     0.2397 & -0.009707 & 0.005482 & -63.545181 \\
     0.5035 & -0.014737 & 0.006620 & -74.456288 \\
     0.7672 & -0.016340 & 0.006510 & -75.869444 \\
     1.0309 & -0.016566 & 0.006208 & -75.973944 \\
     1.2946 & -0.015416 & 0.005953 & -75.865638 \\
     1.5583 & -0.013711 & 0.005811 & -75.738337 \\
     1.8220 & -0.012270 & 0.005758 & -75.624465 \\
     2.0857 & -0.011754 & 0.005843 & -75.528585 \\
     2.3494 & -0.010643 & 0.005758 & -75.449977 \\
     2.6132 & -0.010370 & 0.005778 & -75.386652 \\
     \hline
 \end{tabular}
 \caption{RHF electrostatic, exchange energies for the water dimer system as a function of the OH bond length in monomer 2 and the total RHF energy of monomer 2.\label{tab:rhf_water}}
\end{table}

\begin{table}[h!]
    \begin{tabular}{rrr}
       \hline
       $R$ &     $E_{\mathrm{elst}}$ &    $E_{\mathrm{exch}}$ \\
       \hline
       0.239739 & -0.009756 & 0.005559 \\
       0.503451 & -0.014742 & 0.006671 \\
       0.767163 & -0.016210 & 0.006515 \\
       1.030876 & -0.015636 & 0.006245 \\
       1.294588 & -0.012944 & 0.006107 \\
       1.558300 & -0.009807 & 0.006103 \\
       1.822013 & -0.007061 & 0.006195 \\
       2.085725 & -0.005088 & 0.006305 \\
       2.349438 & -0.004132 & 0.006369 \\
       2.613150 & -0.003750 & 0.006391 \\
       \hline
   \end{tabular}
   \caption{CASCI electrostatic and exchange energies for the water dimer system as a function of the OH bond length in monomer 2. \label{tab:casic_water}}
\end{table}
\begin{table}[h!]
    \begin{tabular}{rrrrrr}
         \hline
         $R$ &  $k$ &     $E_{\mathrm{elst}}$ &    $E_{\mathrm{exch}}$ &       $E_{\mathrm{VQE}}$ &    $\Delta E$ \\
         \hline
         0.2397 &  1 & -0.009756 & 0.005559 & -63.549642 & -0.000029 \\
         0.5035 &  1 & -0.014745 & 0.006672 & -74.463219 & -0.000073 \\
         0.7672 &  1 & -0.016247 & 0.006524 & -75.873483 & -0.000181 \\
         1.0309 &  1 & -0.015858 & 0.006244 & -75.993030 & -0.001175 \\
         1.2946 &  1 & -0.013001 & 0.006113 & -75.921709 & -0.000478 \\
         1.5583 &  1 & -0.009738 & 0.006102 & -75.840780 & -0.005728 \\
         1.8220 &  1 & -0.007066 & 0.006188 & -75.785824 & -0.006303 \\
         2.0857 &  1 & -0.005266 & 0.006288 & -75.752021 & -0.009017 \\
         2.3494 &  1 & -0.004303 & 0.006348 & -75.736872 & -0.012412 \\
         2.6132 &  1 & -0.003858 & 0.006377 & -75.729624 & -0.015426 \\
         \hline
         0.2397 &  2 & -0.009756 & 0.005556 & -63.549658 & -0.000013 \\
         0.5035 &  2 & -0.014737 & 0.006671 & -74.463238 & -0.000054 \\
         0.7672 &  2 & -0.016206 & 0.006514 & -75.873605 & -0.000059 \\
         1.0309 &  2 & -0.015625 & 0.006240 & -75.993936 & -0.000268 \\
         1.2946 &  2 & -0.013014 & 0.006107 & -75.921894 & -0.000294 \\
         1.5583 &  2 & -0.009835 & 0.006102 & -75.846398 & -0.000111 \\
         1.8220 &  2 & -0.007072 & 0.006192 & -75.792062 & -0.000065 \\
         2.0857 &  2 & -0.005091 & 0.006305 & -75.760956 & -0.000082 \\
         2.3494 &  2 & -0.004135 & 0.006369 & -75.749237 & -0.000047 \\
         2.6132 &  2 & -0.003747 & 0.006390 & -75.745000 & -0.000050 \\
         \hline
         0.2397 &  4 & -0.009756 & 0.005560 & -63.549668 & -0.000003 \\
         0.5035 &  4 & -0.014742 & 0.006670 & -74.463277 & -0.000015 \\
         0.7672 &  4 & -0.016210 & 0.006515 & -75.873653 & -0.000011 \\
         1.0309 &  4 & -0.015638 & 0.006245 & -75.994190 & -0.000015 \\
         1.2946 &  4 & -0.012945 & 0.006107 & -75.922173 & -0.000014 \\
         1.5583 &  4 & -0.009807 & 0.006103 & -75.846502 & -0.000006 \\
         1.8220 &  4 & -0.007060 & 0.006194 & -75.792118 & -0.000009 \\
         2.0857 &  4 & -0.005090 & 0.006305 & -75.761019 & -0.000019 \\
         2.3494 &  4 & -0.004130 & 0.006368 & -75.749260 & -0.000024 \\
         2.6132 &  4 & -0.003783 & 0.006400 & -75.745017 & -0.000034 \\
         \hline
     \end{tabular}
     \caption{VQE electrostatic, exchange energies for the water dimer system as a function of the OH bond length in monomer 2 for different values of the $k$ parameter in the $k$-uCJ ansatz. Also given is the total VQE energy of monomer 2 and the deviation of VQE from CASCI. \label{tab:vqe_water}}
\end{table}

\begin{figure}[h!]
    \centering
\begin{BVerbatim}
    O  -1.551007  -0.114520   0.000000
    H  -1.934259   0.762503   0.000000
    H  -0.599677   0.040712   0.000000
    --
    O   1.350625   0.111469   0.000000
    H   1.433068  -0.009834  -0.189640
    H   1.433068  -0.009834   0.189640
\end{BVerbatim}
\caption{Coordinates for H$_2$O dimer given with $R_{\mathrm{OH}} = 0.2397$ in the second monomer.\label{xyz:water}}
\end{figure}

\clearpage
\section{Data and information for KDM5A system}

Below we provide further information about the preparation of the KDM5A structure studied in this work as well as raw energies. Note we provide the xyz files for these structures as attached supplemental files instead of reproducing their coordinates in this pdf.

\subsection{Cutout Preparation for KDM5A \label{app:cutout}}

To make the protein tractable, we used a model system of the binding site which was cut out from the full binding domain of KDM5A. Starting from the Xray structure with ligand 5 cocrystallized (pdb code 6bh4), in the molecular modeling program MOE \citep{VilarMOE2008} the following preparation steps were performed: (1) remove all other small molecule ligands outside the binding site, keeping all Xray waters. (2) add missing side chains with the {\em Prepare Structure} module of MOE. (3) Replace the Mn2+ in the Xray structure with Fe3+. (4) bridge a one-amino-acid break in the protein chain by manually adding and minimizing an alanine residue A357 and cap the protein ends of a missing 9-residue loop with an ACE and NME, respectively. (5) add hydrogens with the {\em Protonate3D} module of MOE. (6) with the MOE module {\em Quickprep} perform a series of tethered minimizations with decreasing tether weights of 100, 50, 10, 5 kcal/mol, keeping all atoms further than 8 A from the ligand fixed, using a maximal gradient of 0.5 kcal/mol \AA$^{-2}$, and a flat well potential $\pm$ 0.25 \AA from the initial atom positions. 
With the protein structure thus prepared, we removed all protein and water residues further than 4.5 A away from the ligand and capped the C- and N-termini of the fragmented protein with NME or ACE. This was followed by additional manual pruning by removing amino acid residues, sidechain or  backbone atoms, do not interact with the ligand or the Fe ion. Only three water molecules were retained – two interacting with the Fe ion, and one interacting with the southern nitrogen atom of the 6-membered aromatic ring of the ligand. The resulting protein model system consisted of 163 atoms, 88 of which were heavy atoms.
Based on the Xray structure of ligand 5, we modelled the binding modes of ligands 4, 8, and 9, and 12. For ligands 8 and 9, a tethered minimization was necessary to resolve a few close contacts between ligand and protein; for ligand 12 with its significantly different hydrogen-bond characteristics, we re-ran {\em Protonate 3D} to optimize the hydrogen-bond network around the ligand.

\subsection{Energy Data}

\begin{table}[h!]
    \begin{tabular}{lrrr}
        \hline
        System    &  RHF           &  VQE (1-uCJ)     & CASCI \\
        \hline
        Protein   & -10829.177918  & -10829.204970    & -10829.220470 \\
        Ligand 4  & -598.793891    &  N/A             & N/A \\
        \hline
        Protein   & -10829.175107  & -10829.201205    & -10829.217807 \\
        Ligand 5  & -715.843329    &  N/A             & N/A \\
        \hline
        Protein   & -10829.238106  & -10829.278175    & -10829.282006 \\
        Ligand 8  & -714.693949    &  N/A             & N/A \\
        \hline
        Protein   & -10829.227796  & -10829.253366    & -10829.269886 \\
        Ligand 9  & -714.714317    &  N/A             & N/A \\
        \hline
        Protein   & -10829.182313  & -10829.206071    & -10829.224031 \\
        Ligand 12 & -692.814843    &  N/A             & N/A \\
        \hline
    \end{tabular}
    \caption{RHF, VQE and CASCI total energies for the relaxed protein and ligands considered in this work. Energies are in Hartree atomic units.}
\end{table}

\begin{table}[h!]
\begin{tabular}{lrrrrrr}
    \hline
    Compound &    $E_{\mathrm{elst}}^{\mathrm{RHF}}$ &     $E_{\mathrm{exch}}^{\mathrm{RHF}}$  & $E_{\mathrm{elst}}^{\mathrm{CASCI}}$ &     $E_{\mathrm{exch}}^{\mathrm{CASCI}}$  & $E_{\mathrm{elst}}^{\mathrm{VQE}}$ &     $E_{\mathrm{exch}}^{\mathrm{VQE}}$  \\
    \hline
    4 &      -0.179402 & 0.129074 & -0.180031 & 0.129294 & -0.179813 & 0.129323 \\
    5 &      -0.177596 & 0.134383 & -0.178216 & 0.134567 & -0.178140 & 0.134560 \\
    8 &      -0.170699 & 0.137911 & -0.171302 & 0.138117 & -0.171266 & 0.138075 \\
    9 &      -0.155808 & 0.139853 & -0.157447 & 0.140533 & -0.156897 & 0.140413 \\
    12 &     -0.164345 & 0.138107 & -0.164988 & 0.138316 & -0.164799 & 0.138228 \\
    \hline
\end{tabular}
\caption{RHF, VQE and CASCI first order SAPT contributions for the relaxed protein and ligands considered in this work. Energies are in Hartree atomic units. Note monomer centered basis sets and the $S^2$ approximation for exchange were employed.}
\end{table}

\begin{table}[h!]
\begin{tabular}{l|r|r}
    \hline
    Component & Compound 4 & Compound 5 \\
    \hline
    $E_\mathrm{elst}$ & -204.242 & -208.646 \\
    $E_\mathrm{exch}$ &  203.780 &  226.393 \\
    $E_\mathrm{ind}$  &  -66.821 &  -68.796 \\
    $E_\mathrm{dis}$  &  -99.107 & -138.142 \\
    \hline
    $E^{(1)}(\mathrm{SAPT})$   &   -0.462 & 17.747 \\
    $E^{(2)}(\mathrm{SAPT})$   & -165.928 & -206.938 \\
    \hline
    $E_\mathrm{int}(\mathrm{SAPT})$         & -166.390 & -189.192 \\
    \hline
\end{tabular}
\caption{Breakdown of SAPT(RHF) contributions to total interaction energy. Energies are in milli-Hartree. Note a \emph{dimer} centered basis set and the $S^\infty$ exchange were employed which should be contrasted to other calculations presented in this work. The SAPT(RHF) method employed here is the `SAPT(0.5)' method of Parrish \emph{et al.} which approximates the dispersion energy\citep{Parrish2018}.}
\end{table}

\clearpage
\bibliography{refs.bib}